**Wavelet Space-Scale-Decomposition Analysis of QSO's Lyα**

**Absorption Lines: Spectrum of Density Perturbations**

Jesus Pando and Li-Zhi Fang

Department of Physics, University of Arizona, Tucson, AZ 85721

## Abstract

A method for measuring the spectrum of a density field by a discrete wavelet space-scale decomposition (SSD) has been studied. We show how the power spectrum can effectively be described by the father function coefficients (FFC) of the wavelet SSD. We demonstrate that the features of the spectrum, such as the magnitude, the index of a power law, and the typical scales, can be determined with high precision by the FFC reconstructed spectrum. This method does not require the mean density, which normally is poorly determined. The problem of the complex geometry of observed samples can also be easily solved because the basis are always orthogonal, regardless the geometry of the samples. Using this method, we examine the spectra inferred from Ly$\alpha$ forests of both simulated and real samples. We find that 1.) the magnitude of the 1-D spectra is significantly different from a Poisson process; 2.) the 1-D spectra are flat on scales less than about 5 h$^{-1}$ Mpc, and show a slow increase with the scale in a range larger than 5 h$^{-1}$ Mpc; 3.) the reconstructed 3-D spectra have about the same power as the COBE normalized linear spectrum of the SCDM model on scales less than 40 h$^{-1}$ Mpc, but the larger than the SCDM model on scales larger than 40 h$^{-1}$ Mpc; 4) the magnitudes of high redshift ($z > 2.51$) spectra generally are larger than those of low redshift ($z < 2.51$) results. Points 3) and 4) are probably caused by large geometric biasing on large scales and high redshifts.

## 1. Introduction

We have recently shown that a space-scale decomposition (SSD) analysis based on the discrete wavelet transform is a powerful tool in detecting structures in the spatial distribution of objects in the universe (Pando & Fang 1996, hereafter PF.) The spatial distributions on various scales can be systematically reconstructed from the mother function coefficients (MFC) of a wavelet based SSD. The clusters can then be identified, scale by scale, from these decomposed distributions. Using this method, the clustering and its evolution of QSO's Ly$\alpha$ forest lines has been studied. The distributions of the wavelet identified clusters were found to be an effective statistical measure which can discriminate among models. Measures such as the number density of Ly$\alpha$ absorption lines, line-line correlations, etc. fail to discriminate between models.

We now continue our work in this direction. The central topic in this paper is the power spectrum of the density perturbation. We will study how to determine power spectrum of a density field by the wavelet SSD. The wavelet based measure of the spectrum is, in some sense, similar to the count in cell (CIC) method. CIC detects the variance $\sigma^2$ of density fluctuations in windows of a cubical cell with side $l$ or Gaussian sphere with radius $R_G$. It is believed that the variance in cell $l$ is mainly contributed by the perturbation on scale $\sim 2l$ or $R_G$. Therefore, the variances should be a measure of the power spectrum on scale $l$ (Efstathiou et al 1990, Saunders 1991, Peacock 1991).

An advantage of the CIC spectrum estimator is that the cells are localized. It reduces the uncertainties caused by a poor knowledge of long wavelength perturbations and by the finite size of the observational samples. All spectrum estimators based on the Fourier transform undergo an infrared (long-wavelength) uncertainty. For instance, the classical spectrum estimator, i.e. the Fourier transform of the autocorrelation function (Peebles 1980) depends essentially on a good measure of the mean density of the sample. It is difficult, or sometimes even impossible, to accurately determine the mean density of objects because of the lack of information of the object's distribution on scales larger than the size of the samples being considered. The Fourier transform of weighted galaxy counts is based on the assumption that all density fluctuations are zero outside of the volume considered (Feldman, Kaiser & Peacock 1994.) If the density fluctuation field is a homogeneous random process, the average of Fourier amplitudes over an *ensemble* of the fluctuation fields with finite extent (zero outside) will be the same as that over an ensemble of the fluctuation field of infinite extent (Adler 1981). Unfortunately, no ensemble is available in cosmology. The effect of the finite size of samples can not be eliminated because the Fourier bases are delocalized. On the other hand, in a CIC analysis by a window of limited spatial support, the behavior of the perturbations on scales larger than the size of a sample does not play an important role.

The problem with the CIC statistics is that its basis (windows) functions are not orthogonal. The variances obtained from the decomposition of cells with different scale $l$ are not independent from each other. Hence, one cannot reconstruct the power spectrum by $\sigma^2(l)$. Moreover, the cubical cell is not localized in Fourier space. We will show that this becomes a severe problem in the case where the power law spectrum has a negative index.

Because of the above-mentioned problem, it is theoretically important to study the spectrum by means of a complete, orthogonal, and localized basis. The wavelet SSD is such a tool. The father function coefficients (FFCs) of the discrete wavelet SSD directly described the fluctuations of density fields on each scale. The variance of the FFCs on each scale is similar to the variance of the CIC. In addition, the basis of the discrete wavelet transform are complete and orthogonal, and it is easy to find the relationship between the coefficients of the Fourier and the wavelet expan-sions. Therefore, one can reconstruct the spectrum from the statistics of the mean of the FFCs and the variance (Yamada & Ohkitani 1991). We apply this method to simulation samples and real data of the Ly$\alpha$ forests.

The contents of the paper are arranged as follows: In §2 and in the Appendix, we describe the method, and derive all formulae needed for the determination of the spectrum via discrete wavelet SSD. §3 demonstrates the wavelet SSD spectrum measure, including reconstruction of power law spectrum, determination of typical scales, and illustration of selection effects by the simulation samples of Ly$\alpha$ forests. In §4 we apply the method to real data of Ly$\alpha$ forests; both 1-dimensional (1-D) and 3-dimensional (3-D) spectra are found. §5 contains discussions and conclusions.

## 2. Wavelet SSD and spectrum of density field

### 2.1 Spectrum in Fourier analysis

Without loss of generality, we consider a 1-D density field $\rho(x)$ over a range $0 \leq x \leq L$. It is convenient to use the density contrast defined by

$$\delta(x) = \frac{\rho(x) - \bar{\rho}}{\bar{\rho}} \tag{1}$$

where $\bar{\rho}$ is the mean density in this field. To express $\delta(x)$ as a Fourier expansion, we take the convention

$$\delta(x) = \sum_{n=-\infty}^{\infty} \delta_n e^{i2\pi nx/L} \tag{2}$$

with the coefficients computed by

$$\delta_n = \frac{1}{L} \int_0^L \delta(x) e^{-i2\pi nx/L} dx \tag{3}$$

Parseval theorem for the Fourier expansion (3) is

$$\frac{1}{L} \int_0^L |\delta(x)|^2 dx = \sum_{n=-\infty}^{\infty} |\delta_n|^2, \tag{4}$$

which shows that the perturbations can be decomposed into domains, $n$, by the orthonormal Fourier basis functions. The power spectrum of perturbations on length scale $L/n$ is then defined as

$$P(n) = |\delta_n|^2. \tag{5}$$

## 2.2 Spectrum with respect to wavelet base

To subject the density contrast $\delta(x)$ to a wavelet expansion, we first assume that $\delta(x)$ is an $L$ periodic function defined on space $-\infty < x < \infty$. A wavelet expansion of $\delta(x)$ is given by (Daubechies 1992, see also Appendix A.2)

$$\delta(x) = \sum_{j=0}^{\infty} \sum_{l=-\infty}^{\infty} \tilde{\delta}_{j,l} \psi_{j,l}(x) \tag{6}$$

where $\psi_{j,l}(x)$ is the base function, also called father function, defined as

$$\psi_{j,l}(x) = \left(\frac{2^j}{L}\right)^{1/2} \psi(2^j x/L - l). \tag{7}$$

The real function $\psi(\eta)$ is the basic wavelet which is localized in a range of $\eta = 0$ to $\eta = 1$ with center at $\eta = 1/2$ (Appendix A.1). Eq.(7) shows that the father functions $\psi_{j,l}(x)$ are generated from the basic function $\psi(x/L)$ by a dilation $2^j$ and a translation $l$. Therefore, the father functions $\psi_{j,l}(x)$ have scale $L/2^j$ and are centered at $lL/2^j$. The bases $\psi_{j,l}(x)$ are complete and orthonormal with respect to both indexes $j$ and $l$. Therefore, the FFCs, $\tilde{\delta}_{j,l}$, in eq.(6) are given by

$$\tilde{\delta}_{j,l} = \int_{-\infty}^{\infty} \delta(x) \psi_{j,l}(x) dx. \tag{8}$$

Similar to the Fourier expansion, the Parseval theorem for the expansion (6) is (see Appendix A.2)

$$\frac{1}{L} \int_0^L |\delta(x)|^2 dx = \sum_{j=0}^{\infty} \frac{1}{L} \sum_{l=0}^{2^j-1} |\tilde{\delta}_{j,l}|^2. \tag{9}$$

Comparing eqs.(9) and (4), one can relate the term $\sum_{l=0}^{2^j-1} |\tilde{\delta}_{j,l}|^2/L$ to the power of perturbations on length scale $L/2^j$, and the term $|\tilde{\delta}_{j,l}|^2/L$ to the power of the perturbation on scale $L/2^j$ at position $lL/2^j$. Therefore, the spectrum with respect to wavelet bases can be defined as

$$P_j = \frac{1}{L} \sum_{l=0}^{2^j-1} |\tilde{\delta}_{j,l}|^2. \tag{10}$$

*2.3 Relationship between $\delta_n$ and $\tilde{\delta}_{j,l}$*

Substituting expansion (2) into eq.(8), we have

$$\tilde{\delta}_{j,l} = \sum_{n=-\infty}^{\infty} \delta_n \int_{-\infty}^{\infty} e^{i2\pi nx/L} \psi_{j,l}(x) dx = \sum_{n=-\infty}^{\infty} \delta_n \hat{\psi}_{j,l}(-n) \tag{11}$$

where $\hat{\psi}_{j,l}(n)$ is the Fourier transform of $\psi_{j,l}(x)$, i.e.

$$\hat{\psi}_{j,l}(n) = \int_{-\infty}^{\infty} \psi_{j,l}(x) e^{-i2\pi nx/L} dx. \tag{12}$$

Using eq.(7), one can rewrite eq.(11) as

$$\tilde{\delta}_{j,l} = \sum_{n=-\infty}^{\infty} \left(\frac{2^j}{L}\right)^{1/2} \delta_n \int_{-\infty}^{\infty} e^{i2\pi nx/L} \psi(2^j x/L - l) dx \tag{13}$$

Defining variable $\eta = 2^j x/L - l$, one finds

$$\tilde{\delta}_{j,l} = \sum_{n=-\infty}^{\infty} \left(\frac{2^j}{L}\right)^{-1/2} \delta_n e^{i2\pi nl/2^j} \int_{-\infty}^{\infty} e^{i2\pi n\eta/2^j} \psi(\eta) d\eta \tag{14}$$

or

$$\tilde{\delta}_{j,l} = \sum_{n=-\infty}^{\infty} \left(\frac{2^j}{L}\right)^{-1/2} \delta_n \hat{\psi}(-n/2^j) e^{i2\pi nl/2^j} \tag{15}$$

where $\hat{\psi}(n)$ is the Fourier transform of the basic function $\psi(\eta)$

$$\hat{\psi}(n) = \int_{-\infty}^{\infty} \psi(\eta) e^{-i2\pi n\eta} d\eta. \tag{16}$$

Eq.(15) is the expression of the FFCs in terms of the Fourier amplitudes.

¿From expansions (3) and (6), one can also express the Fourier coefficient, $\delta_n$, in terms of FFCs as (see Appendix A.3)

$$\delta_n = \frac{1}{L} \sum_{j=0}^{\infty} \sum_{l=0}^{2^j-1} \tilde{\delta}_{j,l} \hat{\psi}_{j,l}(n), \qquad n \neq 0 \tag{17}$$

or

$$\delta_n = \sum_{j=0}^{\infty} \sum_{l=0}^{2^j-1} \left( \frac{1}{2^j L} \right)^{1/2} \tilde{\delta}_{j,l} e^{-i2\pi nl/2^j} \hat{\psi}(n/2^j), \qquad n \neq 0. \tag{18}$$

Eqs.(15) and (18) are the basic equations for detecting spectrum by a wavelet SSD.

### 2.4 Detection of spectrum

The father functions, $\psi_{j,l}(x)$, are compactly supported in Fourier space. For instance, the Fourier transform $\hat{\psi}(n)$ of the Battle-Lemarié wavelet, which is constructed with 4th order spline functions, is non-zero only in two symmetric narrow ranges centered, respectively, at $n = +1$ and $-1$ with widths $\Delta n \ll 1$. For the Daubechies 4 (D4) wavelet, $\hat{\psi}(n)$ also have two symmetric peaks with centers at $n = \pm n_p$ and with width $\Delta n_p$. Therefore, the sum over $n$ in eq.(15) should only be taken on two ranges of $(n_p - 0.5\Delta n_p)2^j \leq n \leq (n_p + 0.5\Delta n_p)2^j$ and $-(n_p + 0.5\Delta n_p)2^j \leq n \leq -(n_p - 0.5\Delta n_p)2^j$. Eq.(15) can be approximately rewritten as

$$\tilde{\delta}_{j,l} \simeq \left( \frac{L}{2^j} \right)^{1/2} \times$$
$$[\hat{\psi}(-n_p) \sum_{n=(n_p-0.5\Delta n_p)2^j}^{(n_p+0.5\Delta n_p)2^j} \delta_n e^{i2\pi nl/2^j} + \hat{\psi}(n_p) \sum_{n=-(n_p+0.5\Delta n_p)2^j}^{-(n_p-0.5\Delta n_p)2^j} \delta_n e^{i2\pi nl/2^j}] \tag{19}$$

$$= \left( \frac{L}{2^j} \right)^{1/2} \sum_{n=(n_p-0.5\Delta n_p)2^j}^{(n_p+0.5\Delta n_p)2^j} [\hat{\psi}(-n_p)\delta_n e^{i2\pi nl/2^j} + \hat{\psi}(n_p)\delta_{-n} e^{-i2\pi nl/2^j}].$$

Eq.(19) shows that the FFCs on scale $j$ are mainly determined by the Fourier components $\delta_n$ with $n$ centered at

$$n = n_p 2^j, \tag{20}$$

where $|n_p|$ are the positions of the peaks of $\psi(n)$.

Because both $\psi(x)$ and $\delta(x)$ are real, we have $\hat{\psi}(-n_p) = \hat{\psi}^*(n_p)$ and $\delta_{-n} = \delta_n^*$. Eq.(19) then becomes

$$
\begin{aligned}
|\tilde{\delta}_{j,l}|^2 &\simeq \frac{L}{2^j} |2 \sum_{n=(n_p-0.5\Delta n_p)2^j}^{(n_p+0.5\Delta n_p)2^j} \mathrm{Re}\{\hat{\psi}(n_p)\delta_n e^{i2\pi n l/2^j}\}|^2 \\
&\simeq \frac{L}{2^j}|\hat{\psi}(n_p)|^2 |2 \sum_{n=(n_p-0.5\Delta n_p)2^j}^{(n_p+0.5\Delta n_p)2^j} \delta_n \cos(\theta_\psi + \theta_n + 2\pi n l/2^j)|^2
\end{aligned}
\tag{21}
$$

where $\theta_\psi$, $\theta_n$ are the phases of $\psi(n_p)$ and $\delta_n$, respectively. In the case of Gaussian perturbations, the distribution of $\theta_n$ is random. Therefore, eq.(21) reduces to

$$
|\tilde{\delta}_{j,l}|^2 \simeq \frac{L}{2^{j-1}} |\hat{\psi}(n_p)|^2 \sum_{n=(n_p-0.5\Delta n_p)2^j}^{(n_p+0.5\Delta n_p)2^j} |\delta_n|^2.
\tag{22}
$$

Eq.(22) shows that the FFCs, $|\tilde{\delta}_{j,l}|^2$, are $l$-independent. This means that different $l$ of $|\tilde{\delta}_{j,l}|^2$ can be treated as independent realizations of stochastic variable of $|\delta_n|^2$ with $n \sim n_p 2^j$. Using the language of the CIC statistic, one can say that each $l$ is a cell, and $\delta_{j,l}$ is the "count" in the cell $l$. Therefore, the square average of FFCs and its variance over the $l$ "realizations" is a measure the spectrum $P(n) = |\delta_n|^2$.

One can use this result to study the conditions under which CIC is a reasonable approach. First, we note that the CIC assumes that the counts in cells with various scales are a scale decomposition. However, this can be guaranteed only if the cells, or window functions, with different scales are orthogonal. The CIC window functions play a role similar to the mother functions in the wavelet SSD. Yet, as we know from discrete wavelet analysis, the mother functions are not orthogonal on different scales. Generally, one cannot find cell functions that form an orthogonal set. Therefore, the MFCs, and then the CIC, are scaled-mixed.

Secondly, the cubic cell CIC corresponds to a window of the form

$$
\phi(\eta) = \begin{cases} 1 & \text{if } 0 \le \eta < 1 \\ 0 & \text{otherwise} \end{cases}
\tag{23}
$$

The corresponding basic function is

$$\psi(\eta) = \begin{cases} 1 & \text{if } 0 \le \eta < 1/2 \\ -1 & \text{if } 1/2 \le \eta < 1 \\ 0 & \text{otherwise} \end{cases} \tag{24}$$

This is just the Haar wavelet [eq.(A3)], whose Fourier transform is given as

$$\hat{\psi}(n) = \frac{2}{\pi n}[\sin(\pi n) - i\cos(\pi n)]\sin^2(\pi n/2) \tag{25}$$

When $n \ll 1$, $\hat{\psi}(n) \sim -i(\pi/2)n$. Therefore, function (24) is not compactly supported in Fourier space. In the case of a power law spectrum, $\delta_n = An^\gamma$, eq.(15) gives

$$\tilde{\delta}_{j,l} = \sum_{n < 2^j} \left(\frac{2^j}{L}\right)^{-1/2} \delta_n \hat{\psi}(-n/2^j) e^{i2\pi nl/2^j} + \text{terms} \ \ n \ge 2^j$$

$$= \sum_{n < 2^j} \left(\frac{2^j}{L}\right)^{-1/2} \frac{i\pi A}{2^{j+1}} n^{1+\gamma} e^{i2\pi nl/2^j} + \text{terms} \ \ n \ge 2^j. \tag{26}$$

Therefore, when $\gamma \le -1$, the large scale (small $n$) perturbations will significantly contribute to, and even dominate, the $\delta_{j,l}$. In this case, the variance $\sigma^2(l)$ of cubical cell CIC on scale $l$ will seriously be contaminated by perturbations on scales larger than $l$, and therefore no longer be a measure of the spectrum at $l$.

## 3. Demonstration of spectrum reconstructions

¿From eq.(10), the spectrum $P_j$ of the wavelet SSD can be detected by the mean square of the FFCs. It can also be measured by the variance of FFCs,

$$P_j^{var} = \frac{1}{L} \sum_{l=0}^{2^j-1} (\overline{\tilde{\delta}_{j,l}} - \tilde{\delta}_{j,l})^2, \tag{27}$$

where $\overline{\tilde{\delta}_{j,l}}$ is the average of $\tilde{\delta}_{j,l}$ over $l$. Because the mean of FFCs, $\overline{\tilde{\delta}_{j,l}}$ is zero [eq.(15)], $P_j$ is equal to $P_j^{var}$. Therefore, the relationship between the spectra in the Fourier analysis and the wavelet SSD is [eq.(22)]

$$P(n)_j \simeq \frac{1}{2^{j+1}\Delta n_p} |\hat{\psi}(n_p)|^{-2} P_j^{var} \tag{28}$$

where $P(n)_j$ is the average of Fourier spectrum on the scale $j$

$$P(n)_j = \frac{1}{2^j \Delta n_p} \sum_{n=(n_p-0.5\Delta n_p)2^j}^{(n_p+0.5\Delta n_p)2^j} P(n), \tag{29}$$

### 3.1 Power index

For a power law spectrum, we have $P(an) = a^\gamma P(n)$, $a$ being any constant, and $\gamma$ the spectrum index. Because eq.(29) now gives $P(n)_{j+1} = 2^\gamma P(n)_j$, eq.(28) shows that

$$P_j \propto 2^{j(\gamma+1)} \tag{30}$$

or

$$\log_2 P_j = (\gamma + 1)j + \text{const} \tag{31}$$

Therefore, the slope of $\log_2 P_j$, when plotted against $j$, is $\gamma + 1$. The index of a power law can be directly found by

$$\gamma = \frac{d \log_2 P_j}{dj} - 1 \tag{32}$$

Figure 1 shows a simple example of a power law spectrum $P(k) = k^{-2}$, where $k \equiv 2\pi n/L$ is the wavenumber in a (length) unit. Using this spectrum we generated distributions over $L$ with bin number $2^9 = 512$, and the bin size is $2\pi$ units. The FFT spectrum reconstruction is plotted in Figure 1a. Figure 1b is the spectra $\log_2 P_j$ and $\log_2 P_j^{var}$ against $j$. The points of $\log_2 P_j^{var}$ in Figure 1b has been shifted down to $\log_2 P_j^{var} - 1$ for presentation purposes only. As expected, Figure 1b shows 1.) $P_j$ is equal to $P_j^{var}$; and 2.) the slopes of the lines $\log_2 P_j - j$ or $\log_2 P_j^{var} - j$ are $-2 + 1 = -1$.

Comparing Figures 1b with 1a one can see that the points of the wavelet spectrum $P(k)_j$ uniformly distribute in $\log_{10} k$ space. Each $P(k)_j$ measures the spectrum $P(k)$ on a scale range of $\log k \rightarrow \log k + \Delta \log k$ and $\Delta \log k = \log 2 = 0.301$. On the other hand, the density of the FFT data in $k$-space is determined by the number of modes, and therefore the distribution of the Fourier spectrum data are uniform with respect to $k$, not $\log k$. The data points are dense at large $\log k$, but rare at small $\log k$.

### 3.2 Normalization factors

¿From eqs.(28) and (20), one can find two useful relations. They are

$$\log P(k)_j = \log P_j - (\log 2)j + A \tag{33}$$

and

$$\log k = (\log 2)j - \log L/2\pi + B \tag{34}$$

Eqs.(33) and (34) transfer the wavelet spectrum $P_j$ or $P_j^{var}$ into the mean Fourier spectrum $P(k)_j$, and *vice versa*. The factor $A$ normalizes the amplitude of $\log P_j$ with $\log P(n)_j$,

$$A = -\log(2\Delta n_p |\hat{\psi}(n_p)|^2) \tag{35}$$

The factor $B$ normalizes the scale $j$ with $\log k$,

$$B = \log n_p. \tag{36}$$

Obviously, the constants $A$ and $B$ depend on the basic wavelet $\psi(\eta)$ being used in the SSD analysis. For instance, in the case of D4 wavelet, $A = 0.602$, and $B = 0.270$.

We tested these normalizations by the following spectrum

$$P(k) = \frac{k}{1 + 10^5 k^4}. \tag{37}$$

This spectrum has a peak at $\log k \sim -1.37$, or a typical scale at $1/k = 23.4$ (length) units. Using (37), we produced samples of distributions over $L$ with bin numbers of 256, 512 and 1024, and the bin size is $2\pi$ units. The reconstruction of spectrum (37) is shown in Figure 2. The peak and the amplitude of the power spectrum are perfectly detected by the wavelet SSD. Therefore, the statistics of $P_j$ or $P_j^{var}$ can effectively provide information of the shape of the spectrum as well as its amplitude.

*3.3 Simulation samples of Lyα forests*

Now we examine the spectrum of the samples given by a simulation of Lyα forests (Bi, Ge & Fang 1995, hereafter BGF). These samples have also been used for the demonstration of cluster identification by the wavelet SSD (Pando & Fang 1996). The simulation was done by the following procedures: 1) generate dark matter distributions by Gaussian perturbations with linear power spectrum of the standard cold dark matter model (SCDM), the cold plus hot dark matter model (CHDM), and the low-density flat cold dark matter model (LCDM); 2) generate the baryonic matter distribution by assuming that baryonic matter traces the dark matter distribution on scales larger than the Jeans length of the baryonic gas, but is smooth over structures on scales less than the Jeans length; 3) remove collapsed regions from the density field because Lyα clouds are probably not virialized; 4) simulate Lyα absorption spectrum as the absorption of neutral hydrogen in the baryonic gas, and include the effects of the observational instrumental point-spread-function, and along with Poisson and background noises; 5) determine the Lyα absorption line and its width from the simulated spectrum by the usual way of Lyα line identification.

Obviously, the samples of the simulated Lyα forests do not only depend on the theoretical spectrum mentioned in step 1), but depend also on the selection effects referred to in steps 2) - 5). One can expect that the reconstructed spectrum will not completely match the theoretical spectra because the simulated spectra are distorted by these selection effects.

The simulated samples cover a redshift range of 1.7 to 4.1. However, in order to consider the redshift-dependence of the spectrum, the samples are synthesized from 13 redshift bins, each with $\Delta z = 0.2$ and centered at $z_n = n \times 0.2 + 1.3$ with $n = 1$ to 13. For redshift bin $n$, the spectrum is taken at redshift $z_n$. Therefore, one cannot use these samples to reconstruct spectrum on scales larger than redshift range of $\Delta z = 0.2$.

We subjected these samples to a SSD spectrum analysis by the D4 wavelet. The 3-D spectrum $P_3(k)$ can be determined from 1-D spectrum $P(k)$ by (see BGF)

$$P(k) = 2\pi \int_k^\infty P_3(q) q \, dq \ , \tag{38}$$

In deriving eq.(38), we have assumed that the random field is statistically isotropic. If the 1-D spectrum can be approximated as a power law $P(k) \propto k^{-\alpha}$, and $\alpha > 0$, the 3-D spectrum is given by

$$\log P_3(k) = \log P(k) - 2 \log k + \log(\alpha/2\pi) \tag{39}$$

Using eq.(34), the wavenumber now is related to $j$ by

$$k = 1.86 \cdot 2\pi \frac{2^j}{D} \ \ \text{h Mpc}^{-1} \tag{40}$$

where $D$ is spatial range of the samples in units of $\text{h}^{-1}$ Mpc. From eqs.(39) nd (40) one can compute 3-D spectrum $P_3(k)$ from 1-D reconstructed spectrum $P(k)_j$.

The reconstructed 3-D spectrum of the SCDM is shown in Figure 3, and the CHDM in Figure 4. The data and 1 $\sigma$ error bar are found from the average over 20 samples in each model. The linear spectra of the SCDM and CHDM used for the simulation are plotted as a gray band in Figures 3 and 4, respectively. The center line of the gray bands is the power spectrum at $z = 2.8$, and the lower and upper edges are the spectra at $z = 4.1$ and $1.7$, respectively.

In the case of the SCDM (Figure 3) the reconstructed spectrum generally agrees with the theoretical spectrum with the difference that power of the reconstructed spectrum shows a faster increase than the model's spectrum when the scale increases.

This is partially because the larger scale structures form later, and small structures earlier. In our first paper (PF) we found that the number density of the larger scale clusters of Ly$\alpha$ lines increases as the redshift decreases. Therefore, the formation of small scale structures is mainly determined by the spectrum at the larger redshifts (the lower part of gray band), while the larger scale structures are determined by the lower redshift (or the upper part of the gray band.)

¿From Figures 3 and 4 one can see that the power of the reconstructed spectrum of the CHDM is lower than the corresponding SCDM. This is expected because the power of the CHDM spectrum is lower than the SCDM.

It is more interesting to note that the power of the reconstructed CHDM spectrum is significantly larger than the theoretical spectrum. This is likely due to the bias of the selections. Because the clustering of CHDM is weak at high redshifts, the clouds identified as Ly$\alpha$ absorbers are rare events, i.e. only the relatively high peaks in the density field are selected. On the other hand, the high peaks in the simulations of the SCDM model have been removed (step 3), therefore the Ly$\alpha$ clouds contain relatively few high peaks. As a consequence of selecting the high peaks, geometric biasing should be considered. The bias effect can also be seen from Figure 3b, which shows that the power of SCDM sample without width selection is lower than the $W > 0.16\mathring{A}$ lines, since the latter is more rare than the former.

The reconstruction at largest scale ($\log k = -0.75$) contains more uncertainty due to the approximation of eq.(39), in which we describe the entire 1-D spectrum $P(k)_j$ by a power law. Actually, the reconstructed 1-D spectrum $P(k)_j$ should not be described by a power law spectrum with the same index on both small and large scales. Because the lack of data on the larger scales, one cannot calculate $P_3(k)$ by the exact formula (38).

## 4. Spectrum determined by Ly$\alpha$ forests

As in our first paper (PF), we look at two data sets of the Ly$\alpha$ forests. The first was compiled by Lu, Wolfe and Turnshek (1991, hereafter LWT). It contains $\sim$ 950 lines from the spectra of 38 QSOs that exhibit neither broad absorption line nor metal line systems. The second is from Bechtold (1994, hereafter JB), which contains a total $\sim$ 2800 lines from 78 QSO's spectra, in which 34 high redshift QSOs were observed at moderate resolution. In our statistics, the effect of proximity to $z_{em}$ has been considered. All lines with redshift $z_{em} \geq z \geq z_{em} - 0.15$ were deleted from our samples. We assumed $q_0 = 1/2$, so the samples covers a comoving distance from about $D_{min}$=2,300 $h^{-1}$Mpc to $D_{max}$=3,300 $h^{-1}$Mpc.

A problem in using real data to do statistics is the complex geometry of QSO's Ly$\alpha$ forests. Different forest covers different spatial ranges, and no one of forests distributes on the entire range of $(D_{min}, D_{max})$. This is a difficulty in detecting the spectrum by usual methods. At the very least, a complicated weighting scheme is needed. However, this problem can easily be solved in the analysis of the wavelet SSD. For a forest sample in a range $(D_1, D_2)$, we can make it to be a sample in the range $(D_{min}, D_{max})$ by adding zero to the data in ranges $(D_{min}, D_1)$ and $(D_2, D_{max})$. Since wavelets are local, the FFCs in the range $(D_1, D_2)$ will not be affected by the addition of zero in the ranges of $(D_{min}, D_1)$ and $(D_2, D_{max})$. We can then compute any statistic by simply dropping all FFCs, $\tilde{\psi}_{j,l}$, with coordinates $l$ in the added zero ranges. Using this technique, geometric complicated samples can be regularized. Therefore all QSO samples can be treated uniformly, and no geometric weighting is needed.

### 4.1 1-D spectrum

The 1-D spectra determined from Ly$\alpha$ forests are shown in Figures 5, 6 and 7. The spectrum are calculated from the LWT data sets with line width $W > 0.36\mathring{A}$, and from JB data set with $W > 0.32\mathring{A}$ and $W > 0.16\mathring{A}$. For each data set, we computed three spectra: a) the entire redshift range $1.7 < z < 4.1$, b) redshift $z > 2.51$ and c) $z < 2.51$. The error bars come from the average over the samples of QSO's absorption spectrum.

The data is such that the constructed spectra are uncertain by a factor of about 10. Nevertheless, some interesting results can already be recognized. The spectra obtained from the two independent data sets, the $W > 0.36\mathring{A}$ LWT and the $W > 0.32\mathring{A}$ JB, show almost the same features, including the magnitudes, the $k$-dependence and the $z$-dependence (see Figures 5 and 6). Therefore, it would be reasonable to consider these features as common properties of the spectra of Ly$\alpha$ forests.

The order of magnitude of the spectra shown in Figures 5 - 7 is $P(k) \sim 0.3 - 10$ h$^{-1}$ Mpc. In computing the spectra for the entire redshift range the finest scale is taken to be $J = 9$, i.e. the number of bins is $N = 2^9 = 512$. For the spectra of the divided redshift ranges, $J = 8$, and $N = 256$. As expected, the magnitudes are not affected by the selection of $J$.

One can compare this magnitude with that of a Poisson process. If the 1-D distribution of the Ly$\alpha$ forests is given by a Poissonian process, the spectrum should be a white noise spectrum and its magnitude is (Vanmarcke 1983)

$$P_{Poisson}(k) = \frac{D}{4\pi N} \quad \text{h}^{-1} \text{ Mpc} \tag{41}$$

where $D$ h$^{-1}$ Mpc is the spatial range of the sample. In our case, $D = D_{max} - D_{min}$ for $J = 9$, and $D = (D_{max} - D_{min})/2$ for $J = 8$, therefore $P_{Poisson} \sim 0.15$ h$^{-1}$ Mpc. This magnitude is less than the mean of $P(k)$ of the real data by a factor of 6. The difference between $P(k)$ and $P_{Poison}(k)$ is larger than $2\sigma$, where $2\sigma$ is the variance of $P(k)$. Therefore, the distribution of Ly$\alpha$ clouds is significantly different from a Poissonian process.

The magnitudes of the $z > 2.51$ spectra generally are larger than those of the $z < 2.51$ results, and the whole redshift range spectra falling in between the two redshift ranges (Figures 5 - 7). Apparently this result conflicts with spectrum evolution: the spectrum magnitude is increasing as the redshift decreases. However, one should not directly identify the redshift-dependence of the reconstructed spectrum magnitudes with the result of the density perturbation evolution because, as discussed in §3.3, the geometric biasing effect has significant impact on the magnitude of the spectrum of the selected objects.

One can also see from Figures 5 - 7 that all the spectra are rather flat on the range of $\log k > 0$, or scales less than about 5 h$^{-1}$ Mpc. This is consistent with the fact that no power of Ly$\alpha$ line-line correlations functions has been detected from these two compiled samples. On scales larger than 5 h$^{-1}$ Mpc, the spectra slightly increase with the increase of scale. However, this increase cannot be detected by two-point correlation function because the correlation function is dominated by noise on such large scales.

Other studies have detected several length scales in the distribution of Ly$\alpha$ forests, including: 1.) 40 h$^{-1}$ Mpc of a void Crotts (1989); 2.) 30-50 h$^{-1}$ Mpc from K-S statistic (Fang 1991); 3.) 80, and even 120 h$^{-1}$ Mpc from typical scale analysis (Mo, et al. 1992a). However, from these results one cannot conclude whether the distribution of Ly$\alpha$ forest lines has multiple typical scales because the different scales may come from the particular method being used. The wavelet SSD uniformly decomposes samples into all scales. Therefore, one can objectively study if the distribution truly has multiple typical scales. Figures 5 - 7 show that no typical scale of 30-50 h$^{-1}$ Mpc exists in the spectrum. Of course, the spectrum detected by the wavelet SSD is the average on range $\Delta \log k = 0.3$, and therefore, will overlook the features of the spectrum with width $\Delta \log k \ll 0.3$.

The only possible typical scale that can be seen in the spectra is 60 - 120 h$^{-1}$ Mpc. Most spectra in Figures 5 - 7 appear to be flat, even dropping at $\log k \sim -1$ to $-1.3$, which is about the same as that given by Mo et al (1992a). This scale has also been claimed at a 90% and higher confidence level by other 1-D sample statistics of pencil-beam redshift surveys of galaxies, QSOs, and CIV absorption line systems in the QSO spectrum (Mo, et al. 1992b; Deng, Xia & Fang 1994).

*4.2 3-D spectrum*

As in §3.3, we constructed the 3-D spectra from the 1-D spectra of the LWT and JB samples. The results are given in Figures 8 and 9. We also plotted the linear spectra of the SCDM and CHDM in Figures 8 and 9. Obviously, one cannot directly test the models by the reconstructed spectrum because the formation of HI clouds underwent non-linear evolution and the identification of Ly$\alpha$ absorption lines underwent selection effects. However, considering that high redshift Ly$\alpha$ clouds are weakly non-linear, it should be interesting to compare constructed spectra with models.

The reconstructed spectra have about the same order of magnitude as the theoretical spectrum of the SCDM model on scales of $\leq 40$ h$^{-1}$ Mpc, but larger than the model on scales $\geq 40$ h$^{-1}$ Mpc (see Figures 8 and 9.) Namely, the differences between the theoretical and constructed spectra increase as the scale increases. Like the systematic differences seen in Figures 3 and 4, these differences may also be due to geometric biasing. It is our experience from cluster identification (PF) that the larger scale clusters of Ly$\alpha$ clouds always locate where the amplitude of the MFCs is higher on smaller scales. Therefore, the larger the scale, the larger the effect of geometric biasing.

The reconstructed 3-D spectra do not show peaks or bending on a scale of about 100 h$^{-1}$ Mpc, which is expected from the models of either SCDM or CHDM. This should not be a surprise. When the scale is less than the bending scale 100 h$^{-1}$ Mpc, one can use eq.(39) to reconstruct the 3-D spectrum from 1-D spectra. However, when the scale is larger than the bending scale, eq.(39) will no longer hold. In this range, $P_3(k) \propto k^\alpha$, with $\alpha > 0$, and eq.(38) shows that 1-D spectrum $P(k)$ is dominated by small scale terms in $P_3(k)$ (Kaiser & Peacock 1991). Therefore, $P(k)$ does not contain information of $P_3(k)$ on scales larger than the bending scale. It is impossible to reconstruct the bending of the 3-D spectrum from 1-D samples.

## 5. Conclusion

We showed that the wavelet SSD is an efficient and reliable tool for detecting the spectrum of density perturbations. For samples of objects tracing the density distribution, the spectrum of density perturbations can be perfectly reconstructed. In this method, no mean density is needed in detecting the spectrum, and the problem of complex geometry of the samples can also be overcome since the wavelet transform bases are always orthogonal and localized, regardless the geometry of the samples. Therefore, the wavelet SSD has great potential in detecting the spectrum of 2-D and 3-D samples.

In this paper, we always followed the convention that the spectrum is given by the decomposition with respect to a Fourier bases. In fact, one can also call $P_j$ or $P_j^{var}$ the spectrum, but with respect to the wavelet bases $\psi_{j,l}$. In principal, the spectrum of density perturbations can be described by any complete and orthogonal bases. The descriptions based on different sets of complete and orthogonal bases should be equivalent if we have an ensemble of realizations of the cosmic stochastic density field. However, no ensemble of realizations is available in cosmology. To overcome this difficulty, it is assumed that the cosmic density field is ergodic. According to this hypothesis, the spectrum can be obtained from the spatial average in one realization of the density distribution (Peebles 1980.) Yet, cosmological observations cannot even provide one realization. Observed samples of large scale distributions must be incomplete as they are constrained by, at the very least, the size of the horizon. Therefore, the information detected by different base sets is not completely equivalent. It can be seen from eqs.(15) and (17) that one cannot transfer the Fourier components $\delta_n$ into wavelet FFCs $\tilde{\delta}_{j,l}$, and vice versa, if the realization is incomplete. Therefore, in terms of large scale structure study, the wavelet SSD $\tilde{\delta}_{j,l}$ is not a mathematical alternative to the Fourier expansion, but instead may reveal some physical features which are missed by $\delta_n$.

Since the basis of the wavelet transform are localized, while the Fourier transform are long-range coherent, the wavelet SSD and Fourier transform measure different aspects of the density fields. The ergodic hypothesis essentially assumes that the spatial correlations are decreasing sufficiently rapidly with increasing separations. The volumes separated with distances larger than the correlation length can be considered as statistically independent regions. Such volumes can then be treated as independent realizations. As we have shown in eq.(22), such independence might easily be described by the FFCs. Therefore, wavelet SSD is probably more effective in picking up information from spatially incomplete samples. This point will become much more clear when we study the non-Gaussianity of the density distributions (Pando & Fang 1995).

Applying wavelet SSD spectrum analysis to samples of the Ly$\alpha$ forests, some common features in their spectra have been revealed. They are: 1.) the magnitude of the 1-D spectra is significantly different from a Poisson process; 2.) the 1-D spectrum are flat on scales less than about 5 h$^{-1}$ Mpc, and very slowly increase with scales larger than 5 h$^{-1}$ Mpc; 3.) the constructed 3-D spectra have about the same power as the linear spectrum of the SCDM model on scales less than 40 h$^{-1}$ Mpc, but larger than the SCDM model on scales larger than 40 h$^{-1}$ Mpc; 4) the magnitudes of high redshift ($z > 2.51$) spectra generally are larger than those of low redshift ($z < 2.51$) results. Points 3) and 4) are probably caused by large geometric biasing on large scales and high redshifts.

Both authors wish to thank Professor P. Carruthers, and Drs.H.G. Bi and P. Lipa

for insightful conversations.

## Appendix

### A.1 Wavelet decomposition of $\delta(x)$

Compactly supported discrete wavelets can be constructed from the scaling function $\phi(\eta)$, which is a solution of the recursive equation (Daubechies 1992, Meyer 1993)

$$\phi(\eta) = \sum_l a_l \phi(2\eta - l) \tag{A1}$$

where $l$ is an integer. If the coefficient $a_l$ are real and satisfy the conditions: $\sum_l a_l a_{l+2m} = 2\delta_{0,m}$ and $\sum_l a_l = 2$, the solution $\phi(x)$ will be orthogonal to integer translates, i.e. one has

$$\int_\infty^\infty \phi(\eta - m)\phi(\eta - l)d\eta = \delta_{m,l} \tag{A2}$$

The basic wavelet $\psi(\eta)$ is defined as

$$\psi(\eta) = \sum_l (-1)^l a_{l+1} \phi(2\eta + l) \tag{A3}$$

The variable $\eta$ is dimensionless.

To conduct a wavelet expansion, one constructs the bases by dilation and translation of $\phi(x)$ and $\psi(x)$ as

$$\phi_m(x) = L^{-1/2}\phi(x/L - m) \tag{A4}$$

and

$$\psi_{j,l}(x) = \left(\frac{2^j}{L}\right)^{1/2} \psi(2^j x/L - l) \tag{A5}$$

where variable $x$ and parameter $L$ have the dimension of length. Father functions $\psi_{j,l}$ and mother functions $\phi_m(x)$, with integer $j$, $l$, $m$ in $-\infty < j, l, m < \infty$ form a complete, orthonormal basis in the space of functions with period length $L$. Therefore, a density distribution $\rho(x)$ can be written as

$$\rho(x) = \sum_{m=-\infty}^{\infty} c_m \phi_m(x) + \bar{\rho} \sum_{j=0}^{\infty} \sum_{l=-\infty}^{\infty} \tilde{\delta}_{j,l}\psi_{j,l}(x) \tag{A6}$$

where $\bar{\rho}$ is the mean density, and the coefficients are calculated by the inner products

$$c_m = \int_{-\infty}^{\infty} \rho(x)\phi_m(x)dx \tag{A7}$$

$$\hat{\delta}_{j,l} = \int_{-\infty}^{\infty} \delta(x)\psi_{j,l}(x)dx \tag{A8}$$

where $\delta(x) = (\rho(x) - \bar{\rho})/\bar{\rho}$ is the density contrast.

By definition, $\rho(x) = \rho(x + mL)$ for integers $m$, eq.(A8) can be rewritten as

$$c_m = \int_{-\infty}^{\infty} \rho(x + mL)\phi_m(x)dx \tag{A9}$$

Using eq.(A4), we have then

$$
\begin{aligned}
c_m &= \int_{-\infty}^{\infty} \rho(x + mL)L^{-1/2}\phi(x/L - m)dx \\
&= \int_{-\infty}^{\infty} \rho(x')L^{-1/2}\phi(x'/L)dx' \\
&= \int_{-\infty}^{\infty} \rho(x')\phi_0(x')dx' = c_0
\end{aligned}
\tag{A10}
$$

where $x' = x + mL$. Therefore coefficients $c_m$ are independent of $m$. Considering the property of "partition of unity" of the scaling function (Daubechies 1992)

$$\sum_{m=-\infty}^{\infty} \phi(\eta + m) = 1, \tag{A11}$$

one has from eq.(A10)

$$
\begin{aligned}
c_0 &= \int_{-\infty}^{\infty} \rho(x)L^{-1/2}\phi(x/L)dx \\
&= \sum_{-\infty}^{\infty} \int_0^L \rho(x + mL)L^{-1/2}\phi(x/L + m)dx \\
&= L^{-1/2} \int_0^L \rho(x)\sum_{-\infty}^{\infty} \phi(x/L + m)dx \\
&= L^{-1/2} \int_0^L \rho(x)dx.
\end{aligned}
\tag{A12}
$$

The first term in the expansion (A6) becomes

$$\sum_{m=-\infty}^{\infty} c_m\phi_m(x) = c_0 L^{-1/2} \sum_{m=-\infty}^{\infty} \phi(x/L + m) = L^{-1} \int_0^L \rho(x)dx. \tag{A13}$$

Therefore, this term is the mean density $\bar{\rho}$. Because all basis functions $\int_{-\infty}^{\infty} \psi_{j,l}(x)dx = 0$, one can find finally from eq.(A6)

$$\delta(x) = \frac{\rho(x) - \bar{\rho}}{\bar{\rho}} = \sum_{j=0}^{\infty} \sum_{l=-\infty}^{\infty} \tilde{\delta}_{j,l} \psi_{j,l}(x) \tag{A14}$$

where the father function coefficients $\tilde{\delta}_{j,l}$ are given by eq.(A8) or eq.(8).

## A.2 Parseval theorem

¿From eq.(8) or (A8), we have

$$
\begin{aligned}
\tilde{\delta}_{j,l+K} &= \int_{-\infty}^{\infty} \delta(x) \psi_{j,l+K}(x) dx \\
&= \int_{-\infty}^{\infty} \delta(x) \left( \frac{2^j}{L} \right)^{1/2} \psi(2^j x/L - l - K) dx \\
&= \int_{-\infty}^{\infty} \delta(x' + 2^{-j} KL) \left( \frac{2^j}{L} \right)^{1/2} \psi(2^j x'/L - l) dx' \tag{A15}
\end{aligned}
$$

here we make a change of variable $x' = x - 2^{-j} KL$. Therefore, when $2^{-j} K$ is an integer, i.e. $K = 2^j m$, eq.(A15) is

$$\tilde{\delta}_{j,l+2^j m} = \int_{-\infty}^{\infty} \delta(x' + mL) \left( \frac{2^j}{L} \right)^{1/2} \psi(2^j x'/L - l) dx' = \tilde{\delta}_{j,l} \tag{A16}$$

which shows that the FFC, $\tilde{\delta}_{j,l}$, are periodic in $l$ with period $2^j$.

We now show the Parseval theorem. From the expansion (A14) we have

$$\int_0^L |\delta(x)|^2 dx = \sum_{j,j'=0}^{\infty} \sum_{l,l'=-\infty}^{\infty} \tilde{\delta}_{j,l} \tilde{\delta}_{j',l'} \int_0^L \psi_{j,l}(x) \psi_{j',l'}(x) dx \tag{A17}$$

Considering the periodicity (A16), eq.(A17) can be rewritten as

$$
\begin{aligned}
&\int_0^L |\delta(x)|^2 dx \\
&= \sum_{j,j'=0}^{\infty} \sum_{l=0}^{2^j-1} \sum_{m=-\infty}^{\infty} \sum_{l'=0}^{2^{j'}-1} \sum_{m'=-\infty}^{\infty} \tilde{\delta}_{j,l} \tilde{\delta}_{j',l'} \int_0^L \psi_{j,l+2^j m}(x) \psi_{j',l'+2^{j'} m'}(x) dx \\
&= \sum_{j,j'=0}^{\infty} \sum_{l=0}^{2^j-1} \sum_{l'=0}^{2^{j'}-1} \tilde{\delta}_{j,l} \tilde{\delta}_{j',l'} \times
\end{aligned}
$$

$$\sum_{m=-\infty}^{\infty} \sum_{m'=-\infty}^{\infty} \frac{2^j}{L} \int_0^L \psi(2^j x/L - l - 2^j m)\psi(2^{j'} x/L - l' - 2^{j'} m')dx$$

$$= \sum_{j,j'=0}^{\infty} \sum_{l=0}^{2^j-1} \sum_{l'=0}^{2^{j'}-1} \tilde{\delta}_{j,l}\tilde{\delta}_{j',l'} \sum_{m"\equiv(m-m')=-\infty}^{\infty} \sum_{m'=-\infty}^{\infty}$$

$$\frac{2^j}{L} \int_0^L \psi(2^j x/L - l - 2^j(m"+m'))\psi(2^{j'} x/L - l' - 2^{j'} m')dx$$

$$= \sum_{j,j'=0}^{\infty} \sum_{l=0}^{2^j-1} \sum_{l'=0}^{2^{j'}-1} \tilde{\delta}_{j,l}\tilde{\delta}_{j',l'} \times$$

$$\sum_{m"=-\infty}^{\infty} \sum_{m'=-\infty}^{\infty} \frac{2^j}{L} \int_0^L \psi(2^j(x/L - m') - l - 2^j m")\psi(2^{j'}(x/L - m') - l')dx$$

$$= \sum_{j,j'=0}^{\infty} \sum_{l=0}^{2^j-1} \sum_{l'=0}^{2^{j'}-1} \tilde{\delta}_{j,l}\tilde{\delta}_{j',l'} \times$$

$$\sum_{m"=-\infty}^{\infty} \frac{2^j}{L} \int_{-\infty}^{\infty} \psi(2^j x/L - l - 2^j m")\psi(2^{j'} x/L - l')dx$$

$$= \sum_{j,j'=0}^{\infty} \sum_{l=0}^{2^j-1} \sum_{l'=0}^{2^{j'}-1} \tilde{\delta}_{j,l}\tilde{\delta}_{j',l'} \sum_{m"=-\infty}^{\infty} \int_{-\infty}^{\infty} \psi_{j,l+2^j m"}(x)\psi_{j',l'}(x)dx$$

$$= \sum_{j,j'=0}^{\infty} \sum_{l=0}^{2^j-1} \sum_{l'=0}^{2^{j'}-1} \tilde{\delta}_{j,l}\tilde{\delta}_{j',l'} \sum_{m"=-\infty}^{\infty} \delta_{j,j'}\delta_{l+2^j m",l'} \qquad (A18)$$

¿From $\delta_{j,j'}$, $j'$ should be equal to $j$, and then $l' < 2^j$. Therefore, $\delta_{l+2^j m",l'}$ requires $m" = 0$ and $l = l'$. We have then the Parseval theorem (9).

## A.3 Derivation of eqs.(17) and (18)

Substituting the wavelet expansion of $\delta(x)$ (eq.(A6)) into eq.(3), we have

$$\delta_n = \frac{1}{L} \int_0^L \left[ \sum_{j=0}^{\infty} \sum_{l=-\infty}^{\infty} \tilde{\delta}_{j,l}\psi_{j,l}(x) \right] e^{-i2\pi nx/L}dx \qquad (A19)$$

Using eq.(A16), eq.(18) becomes

$$\delta_n = \frac{1}{L} \int_0^L \left[ \sum_{j=0}^{\infty} \sum_{l=0}^{2^j-1} \sum_{m=-\infty}^{\infty} \tilde{\delta}_{j,l}\psi_{j,l+2^j m}(x) \right] e^{-i2\pi nx/L}dx$$

$$= \frac{1}{L} \sum_{j=0}^{\infty} \sum_{l=0}^{2^j-1} \tilde{\delta}_{j,l} \sum_{m=-\infty}^{\infty} \int_0^L \left( \frac{2^j}{L} \right)^{1/2} \psi(2^j x/L - l - 2^j m)e^{-i2\pi nx/L}dx$$

$$= \frac{1}{L} \sum_{j=0}^{\infty} \sum_{l=0}^{2^j-1} \tilde{\delta}_{j,l} \int_{-\infty}^{\infty} \left(\frac{2^j}{L}\right)^{1/2} \psi(2^j x'/L - l) e^{-i2\pi n x'/L} dx'$$

$$= \frac{1}{L} \sum_{j=0}^{\infty} \sum_{l=0}^{2^j-1} \tilde{\delta}_{j,l} \int_{-\infty}^{\infty} \psi_{j,l}(x') e^{-i2\pi n x'/L} dx'$$

$$= \frac{1}{L} \sum_{j=0}^{\infty} \sum_{l=0}^{2^j-1} \tilde{\delta}_{j,l} \hat{\psi}_{j,l}(n) \qquad (A20)$$

This eq.(17). An alternative form, which uses the Fourier transform of the basic function $\psi(x)$ rather than $\psi_{j,l}(x)$, can be derived from eq.(A20) as follows

$$\delta_n = \frac{1}{L} \sum_{j=0}^{\infty} \sum_{l=0}^{2^j-1} \tilde{\delta}_{j,l} \int_{-\infty}^{\infty} \psi_{j,l}(x) e^{-i2\pi n x/L} dx$$

$$= \frac{1}{L} \sum_{j=0}^{\infty} \sum_{l=0}^{2^j-1} \tilde{\delta}_{j,l} \int_{-\infty}^{\infty} \left(\frac{2^j}{L}\right)^{1/2} \psi(2^j x/L - l) e^{-i2\pi n x/L} dx$$

$$= \frac{1}{L} \sum_{j=0}^{\infty} \sum_{l=0}^{2^j-1} \left(\frac{2^j}{L}\right)^{-1/2} \tilde{\delta}_{j,l} e^{-i2\pi n l/2^j} \int_{-\infty}^{\infty} \psi(\eta) e^{-i2\pi n \eta/2^j} d\eta$$

$$= \sum_{j=0}^{\infty} \sum_{l=0}^{2^j-1} \left(\frac{1}{2^j L}\right)^{1/2} \tilde{\delta}_{j,l} e^{-i2\pi n l/2^j} \hat{\psi}(n/2^j) \qquad (A21)$$

This is eq.(18).

**Figure captions**

**Figure 1** Reconstruction of a power law spectrum $P(k) = k^{-2}$ by a.) FFT, and b.) wavelet SSD, where $k = 2\pi n/L$ is the wavenumber in a length unit. The points of $\log_2 P^{var}$ (star) in 1b have been shifted down to $\log_2 P_j^{var} - 1$ for clarity of presentation only. The slopes of the lines $\log_2 P_j - j$ (diamond) or $\log_2 P_j^{var} - j$ are $-2 + 1 = -1$.

**Figure 2** Reconstruction of spectrum with a typical scale eq.(37). $k = 2\pi n/L$ is the wavenumber in a length unit. The samples in a, b and c are produced from spectrum (36) on $L$ with the bin numbers of 256, 512 and 1024, and the bin size is $2\pi$ length units. The peak of the spectrum is at $\log k$ = -1.37.

**Figure 3** Reconstruction of 3-D spectrum from simulated Ly$\alpha$ forests samples of the SCDM model: a.) for lines with width larger than 0.16 $\mathring{A}$ and b.) for lines without width selection. The data and 1 $\sigma$ error bar are found from the average over 20 simulated samples. $k$ is in units h Mpc$^{-1}$. The gray band is the linear spectra of the SCDM. The center line of the gray bands is the power spectrum at $z = 2.8$, and the lower and upper edges are the spectra at $z = 4.1$ and 1.7, respectively.

**Figure 4** Reconstruction of 3-D spectrum from simulated Ly$\alpha$ forests samples with width larger than 0.16 $\mathring{A}$ in the CHDM model. The gray band is the linear spectra of the CHDM. The center line of the gray bands is the power spectrum at $z = 2.8$, and the lower and upper edges are the spectra at $z = 4.1$ and 1.7, respectively.

**Figure 5** 1-D spectra $P(k)_j$ (diamond) and $P^{var}(k)_j$ (star) of LWT Ly$\alpha$ forest samples with width > 0.36 $\mathring{A}$. a.) is the spectrum given by data in the entire redshift

range $1.7 < z < 4.1$; b.) for redshift range $z < 2.51$ and c.) for $z > 2.51$. $k$ is in unit h $Mpc^{-1}$. The error bars are obtained from the average over the samples of QSO's absorption spectrum. For clarity, the points $P^{var}(k)_j$ are plotted at $\log k + 0.05$.

**Figure 6** The same as Figure 5 for sample of JB Ly$\alpha$ forest with width $> 0.32$ $\mathring{A}$.

**Figure 7** The same as Figure 5 for sample of JB Ly$\alpha$ forest with width $> 0.16$ $\mathring{A}$.

**Figure 8** Reconstruction of 3-D spectra given by $P(k)_j$ (diamond) and $P^{var}(k)_j$ (star) of LWT Ly$\alpha$ forest samples with width $> 0.36$ $\mathring{A}$. For clarity, the points reconstructed by variance are plotted at $\log k + 0.05$. The gray and dark bands are the linear spectra of the SCDM and CHDM, respectively.

**Figure 9** The same as Figure 8 for a.) $W > 0.32$ $\mathring{A}$ and b.) $W > 0.16$ $\mathring{A}$ of JB Ly$\alpha$ forest samples.

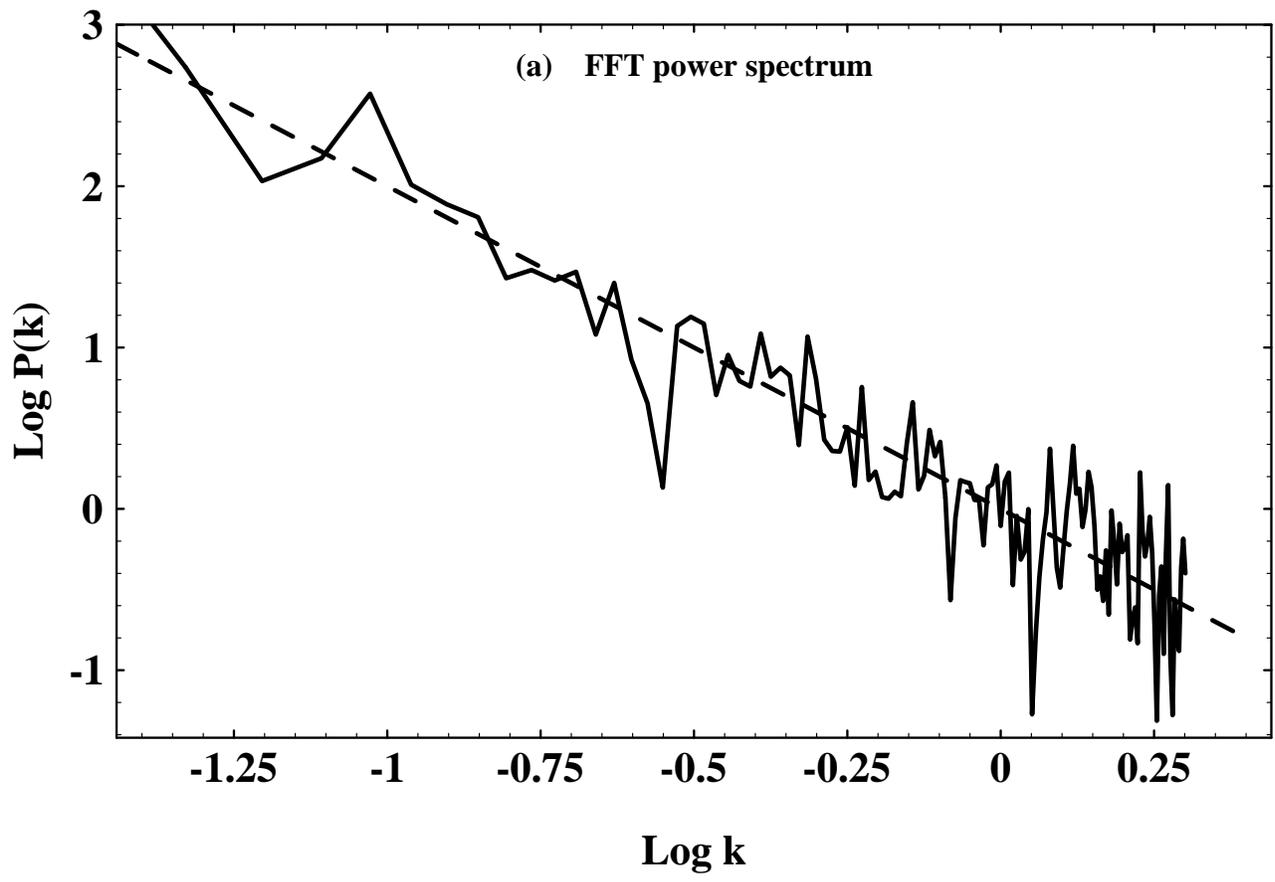

**(a) FFT power spectrum**

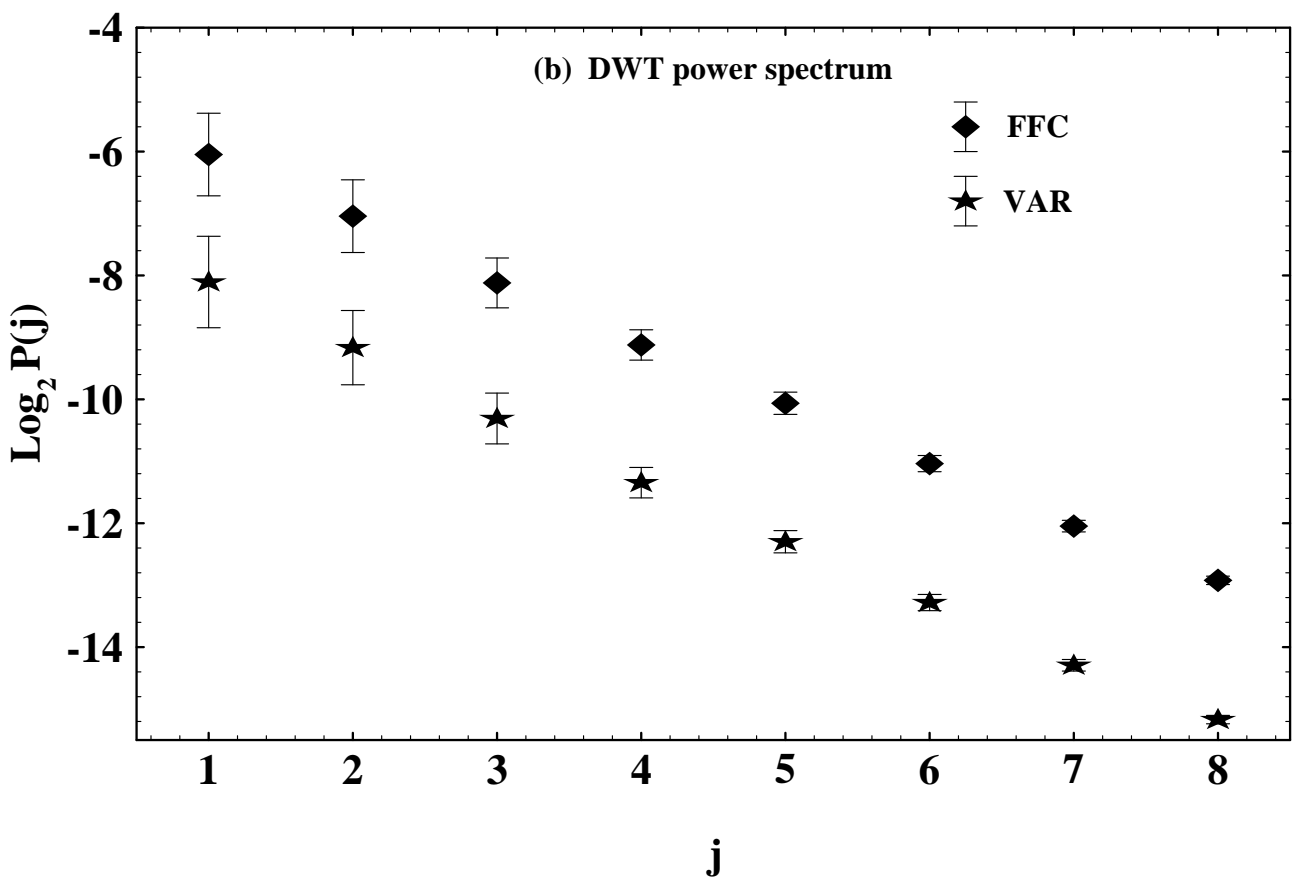

**(b) DWT power spectrum**

FFC

VAR

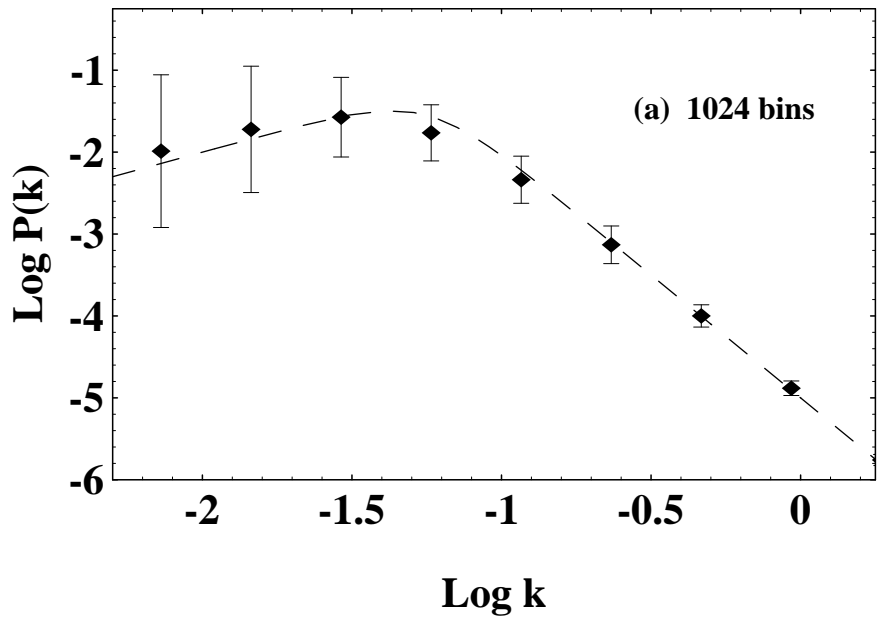

(a) 1024 bins

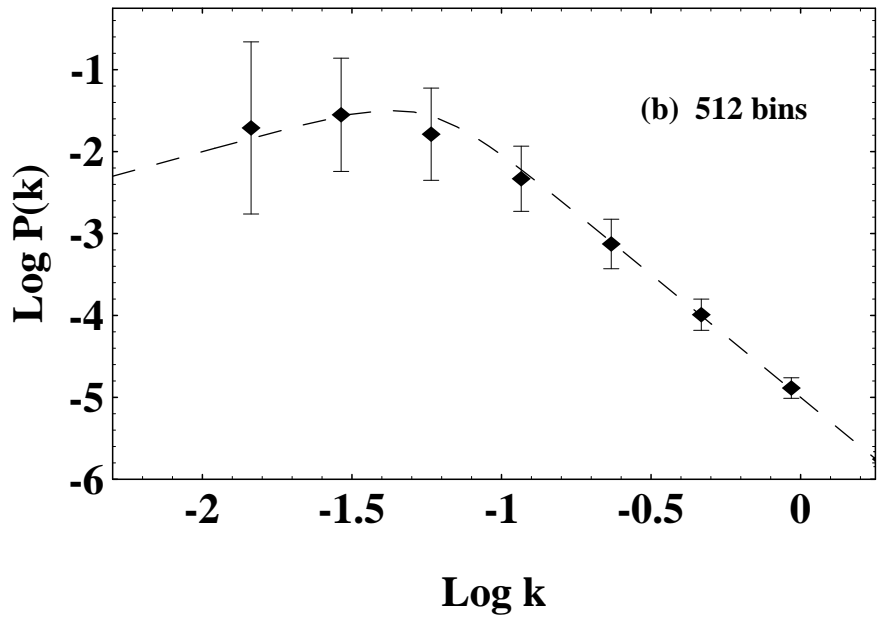

(b) 512 bins

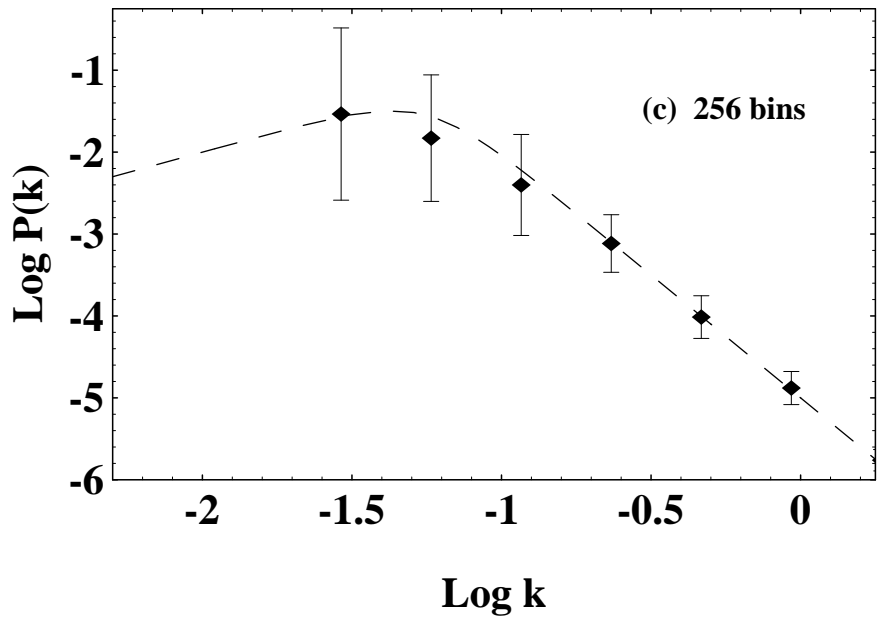

(c) 256 bins

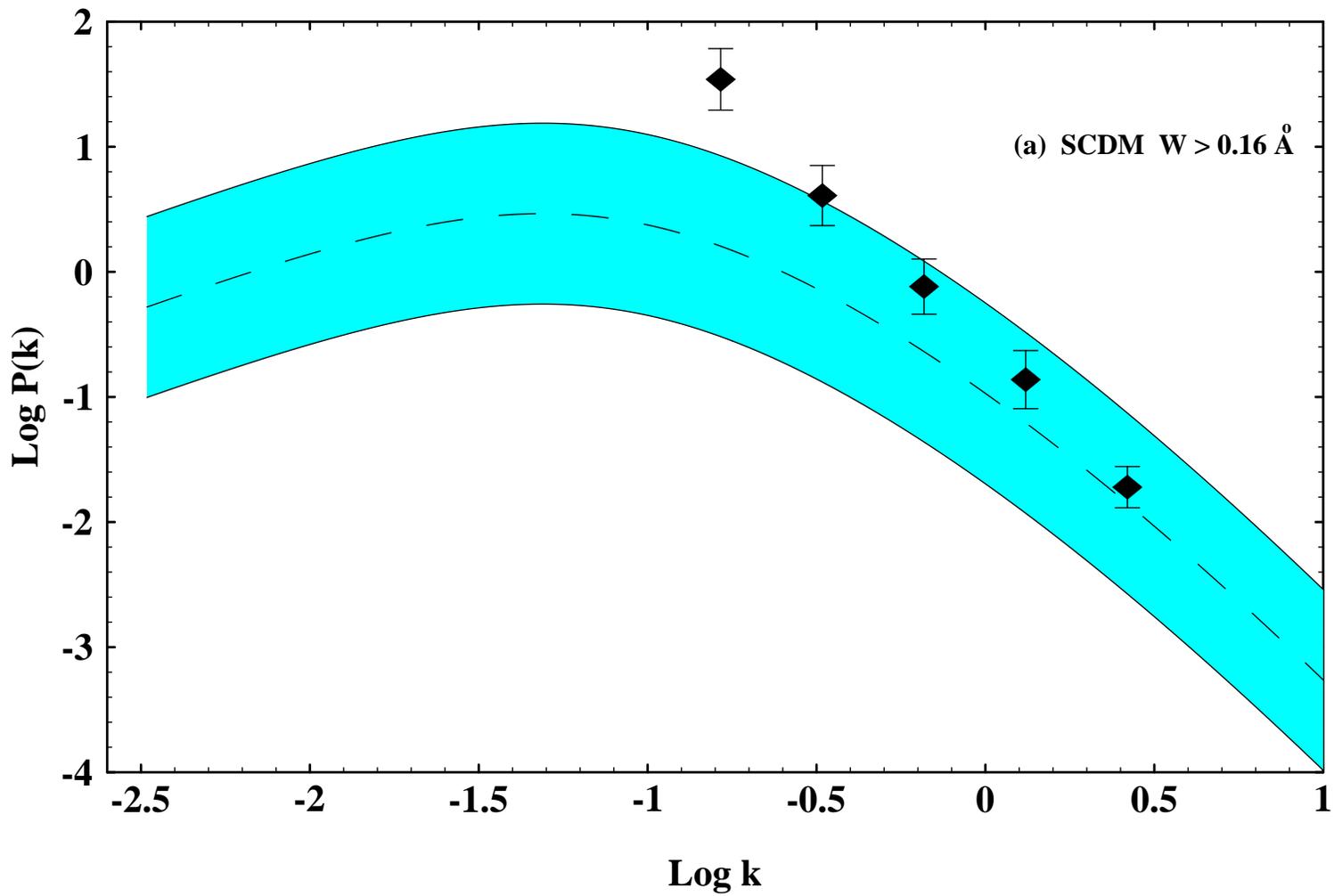

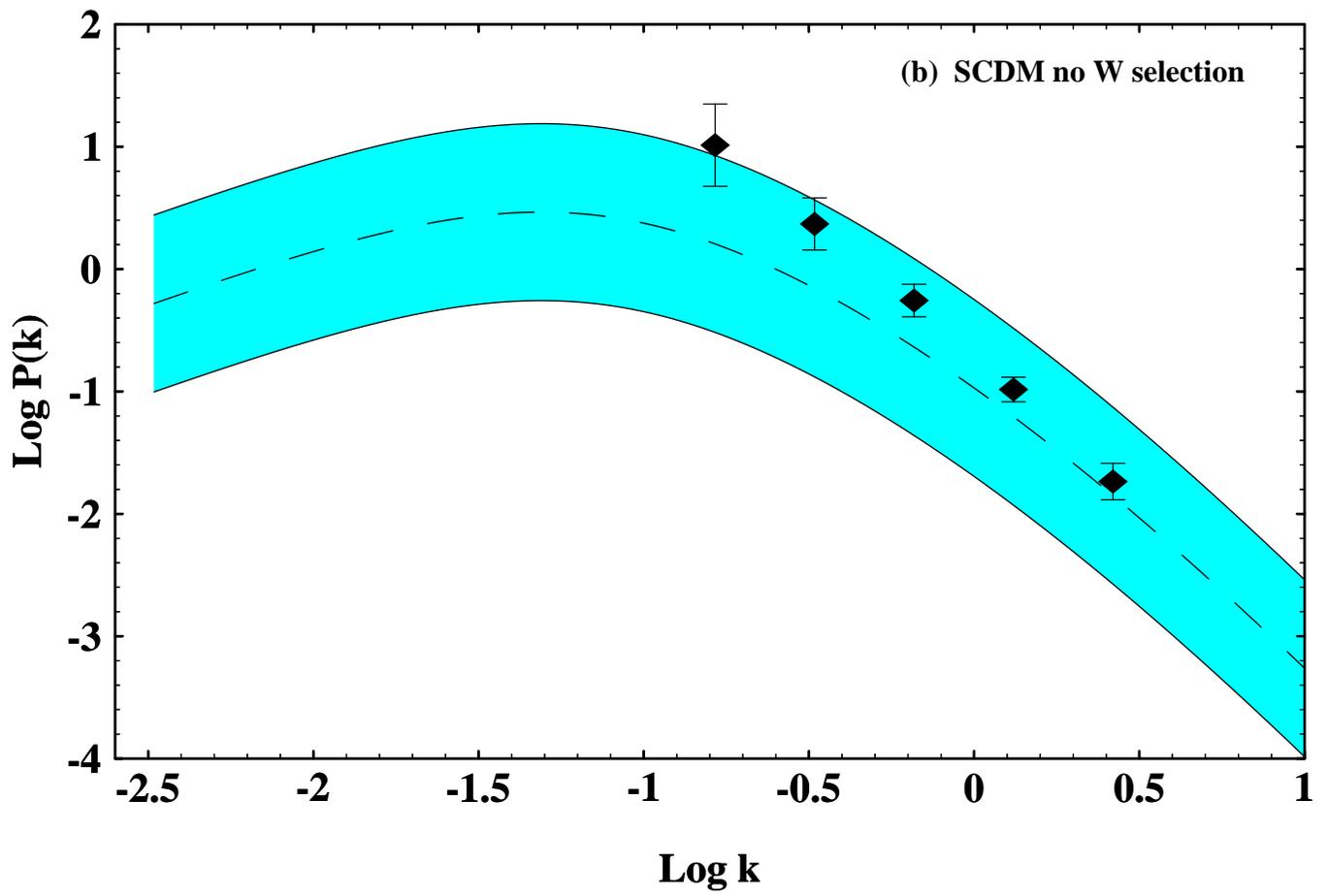

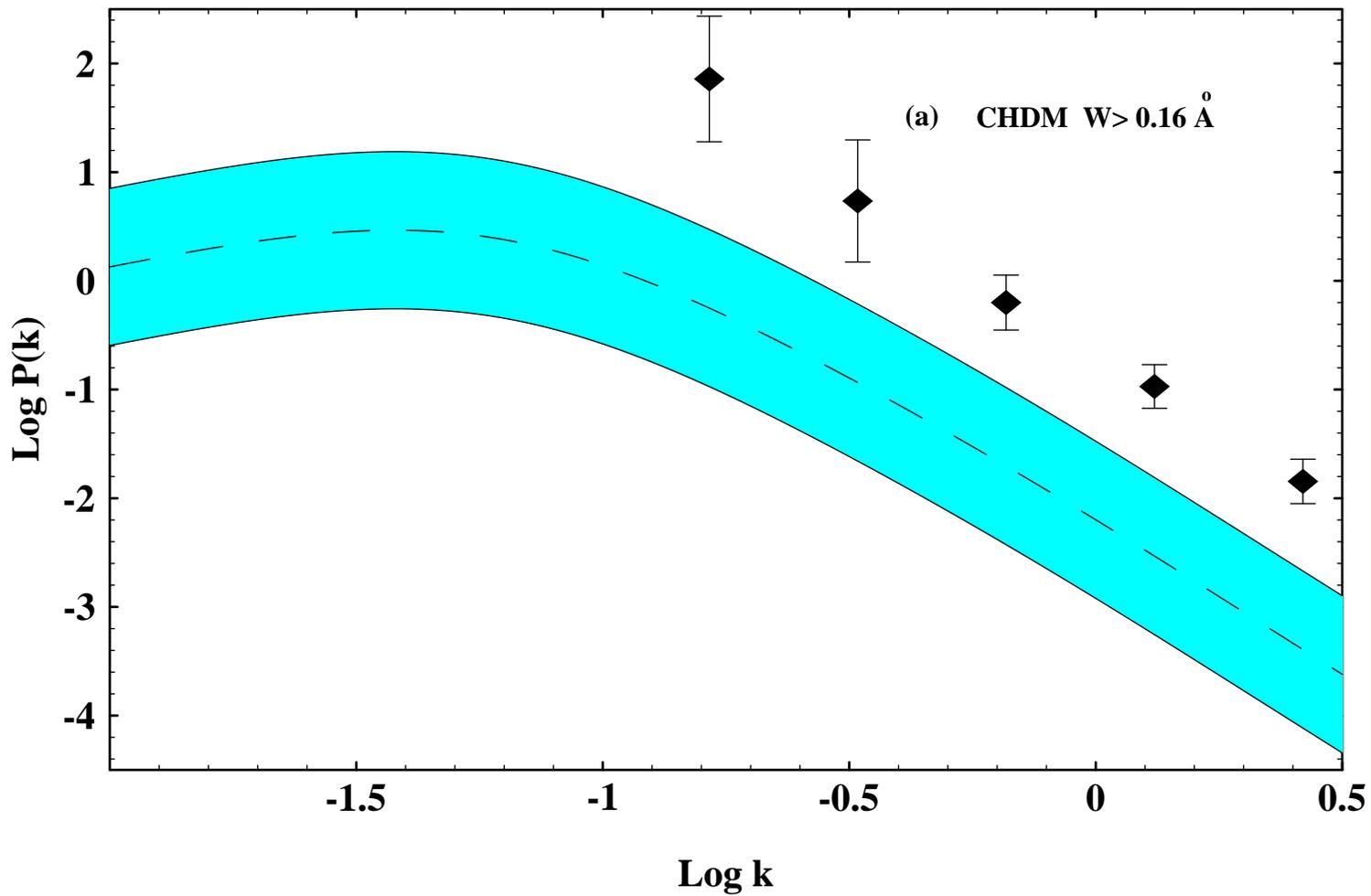

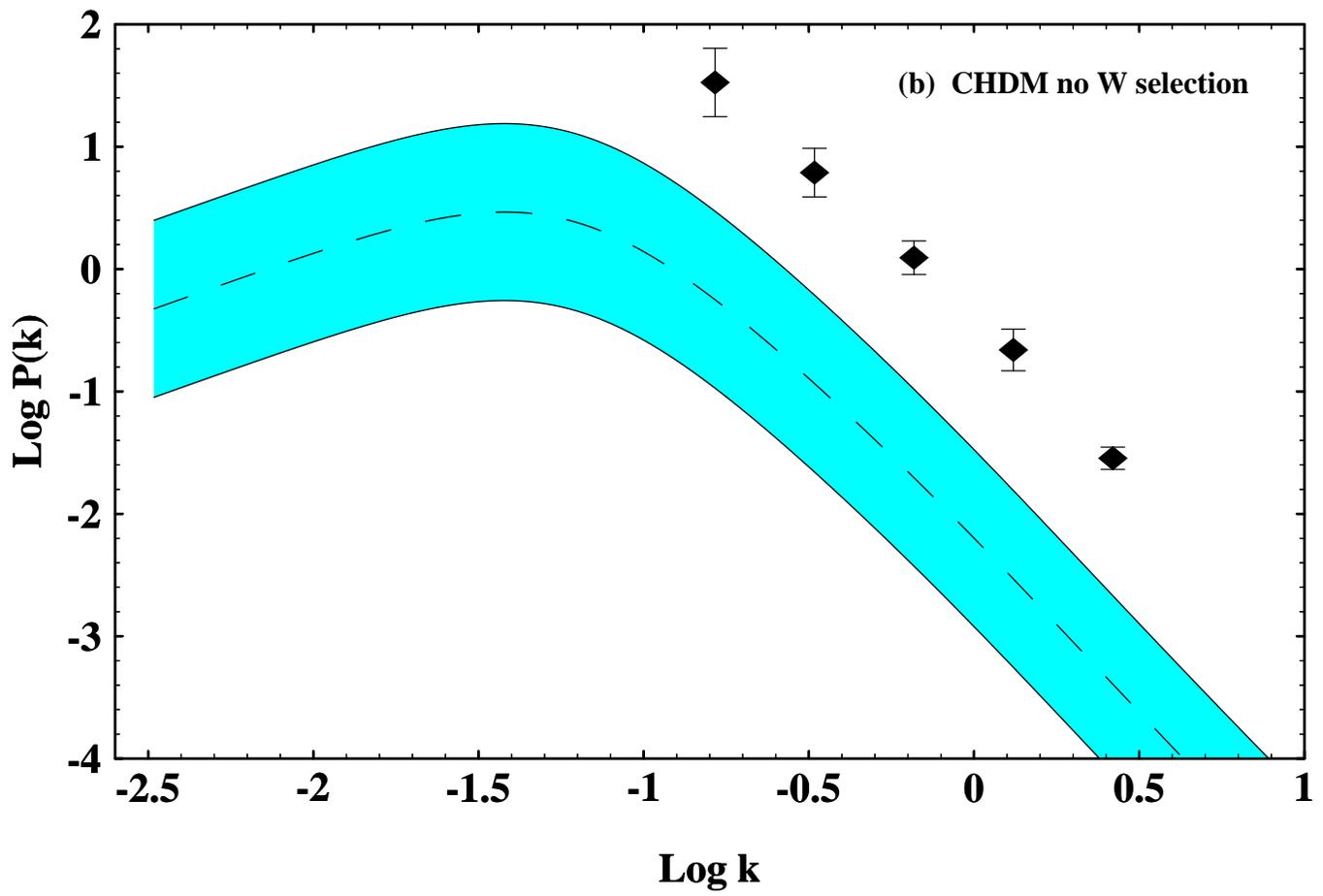

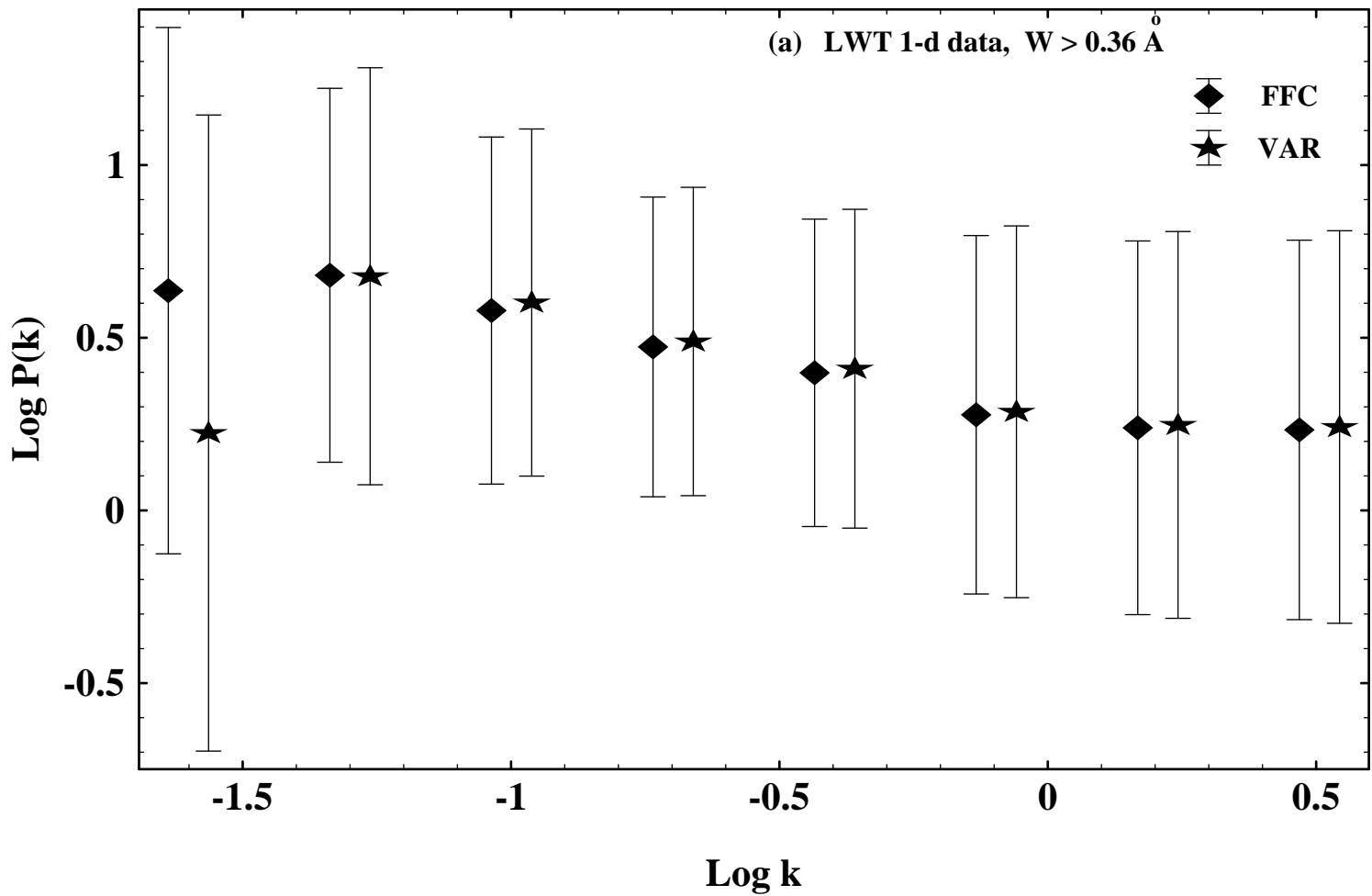

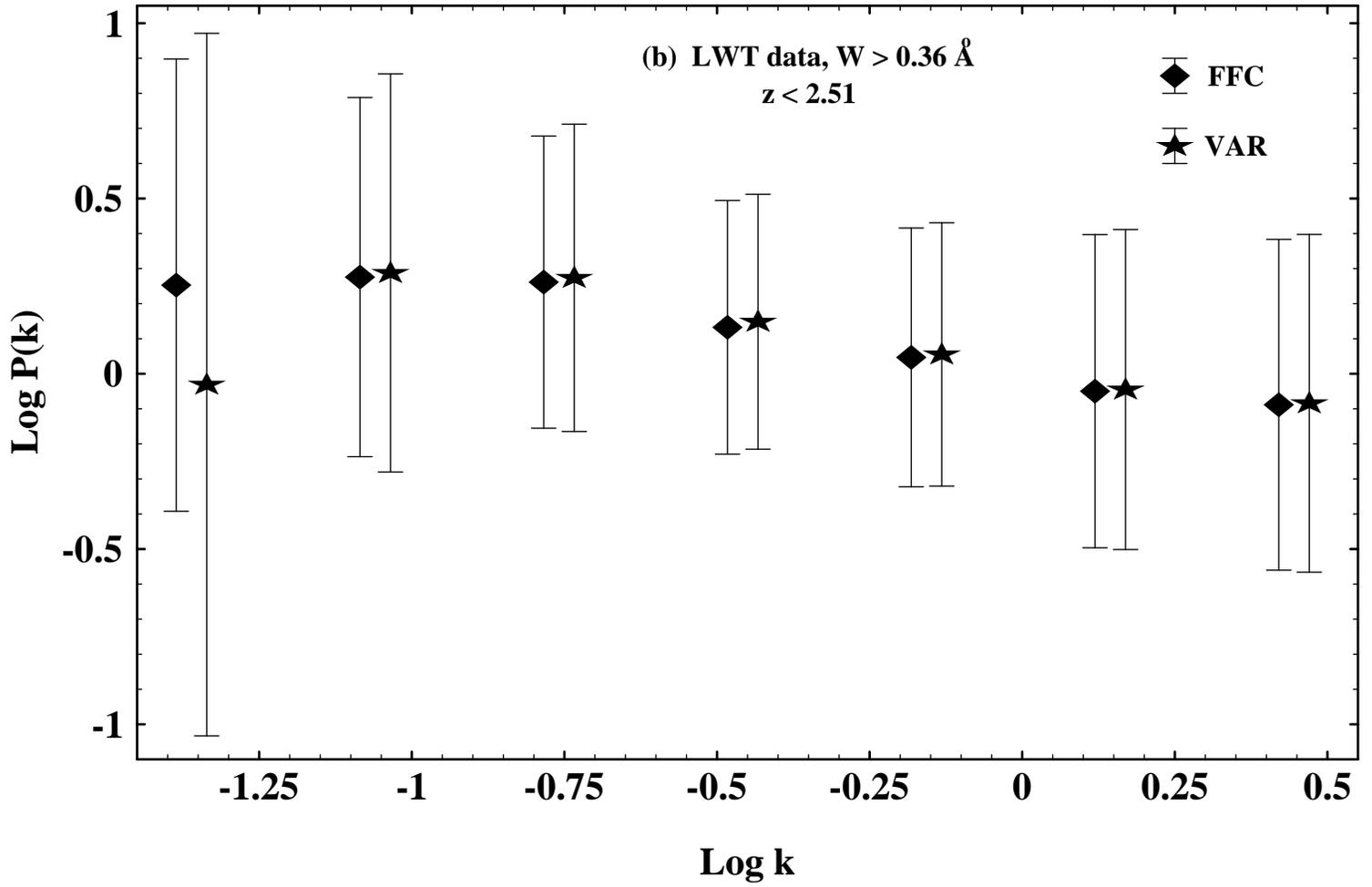

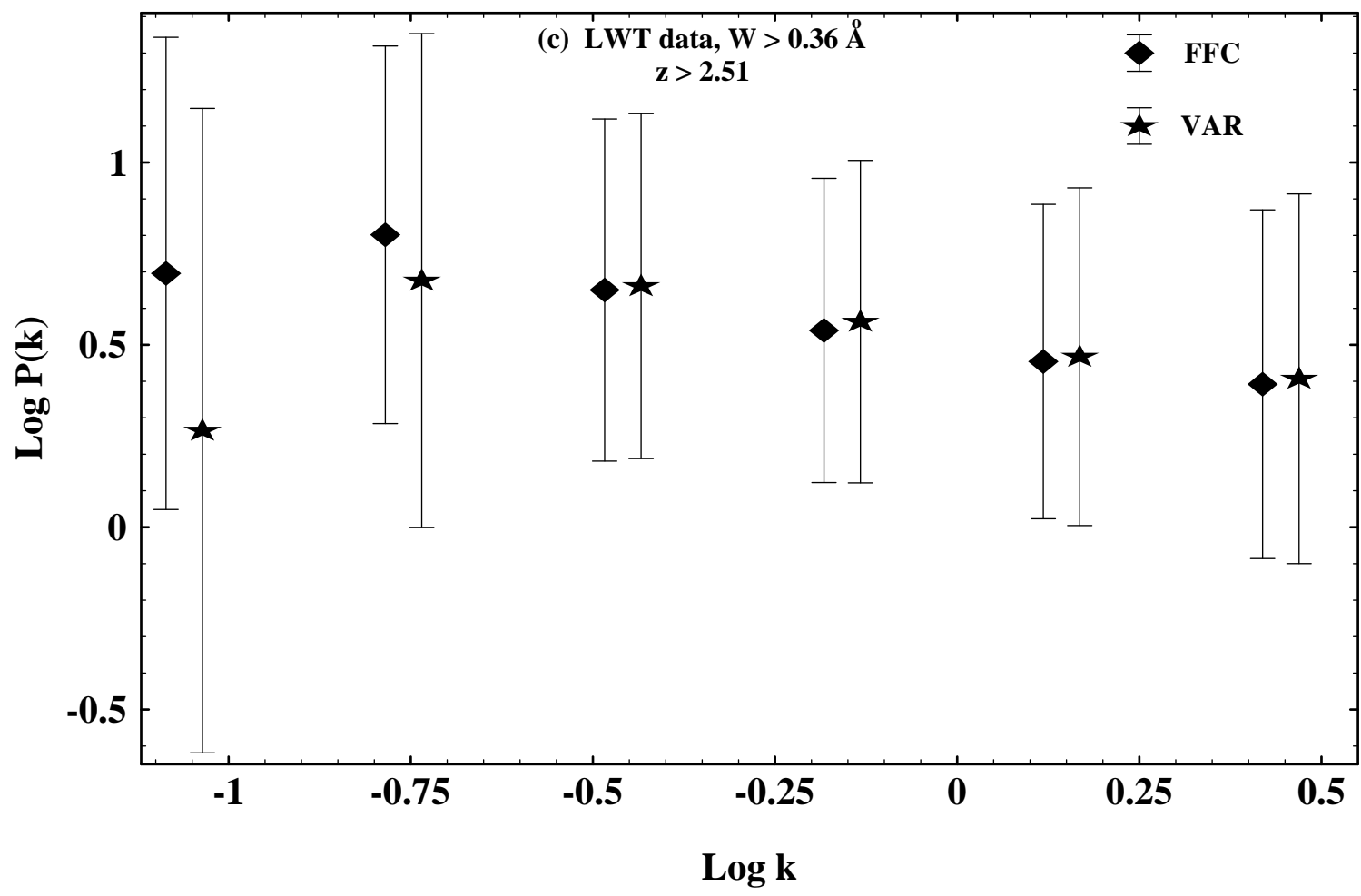

(c) LWT data, W > 0.36 Å
z > 2.51

FFC ◆
VAR ★

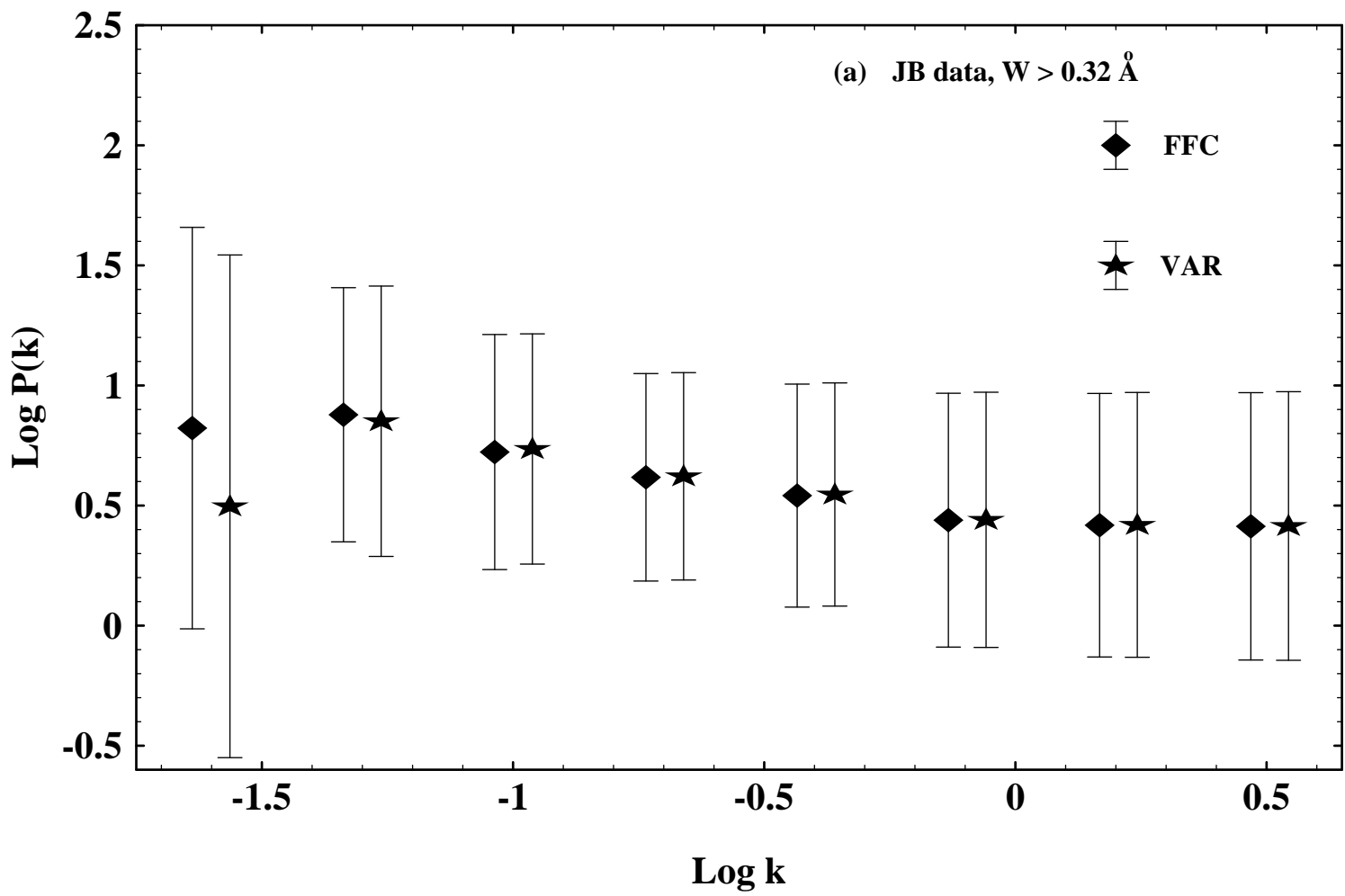

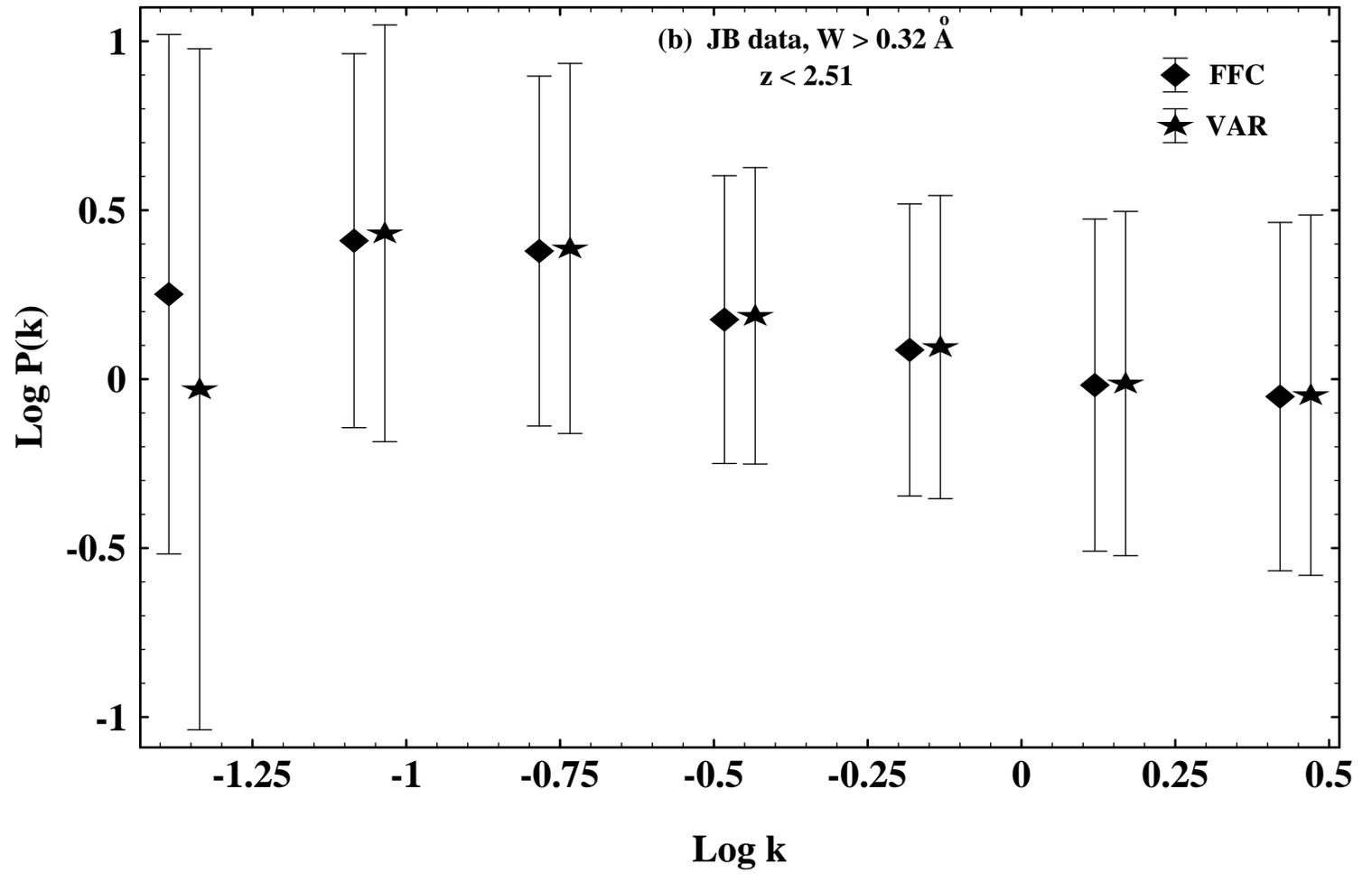

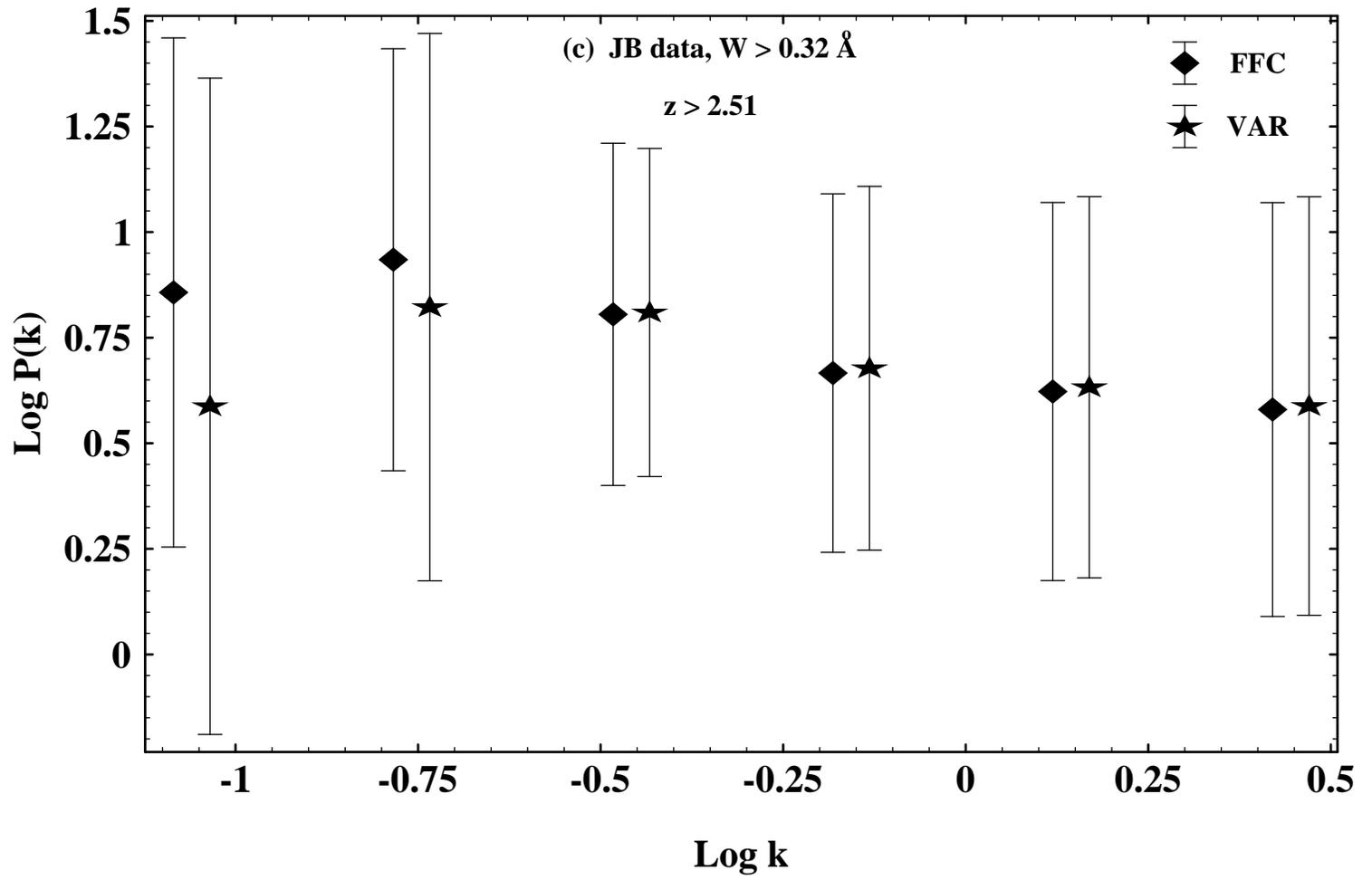

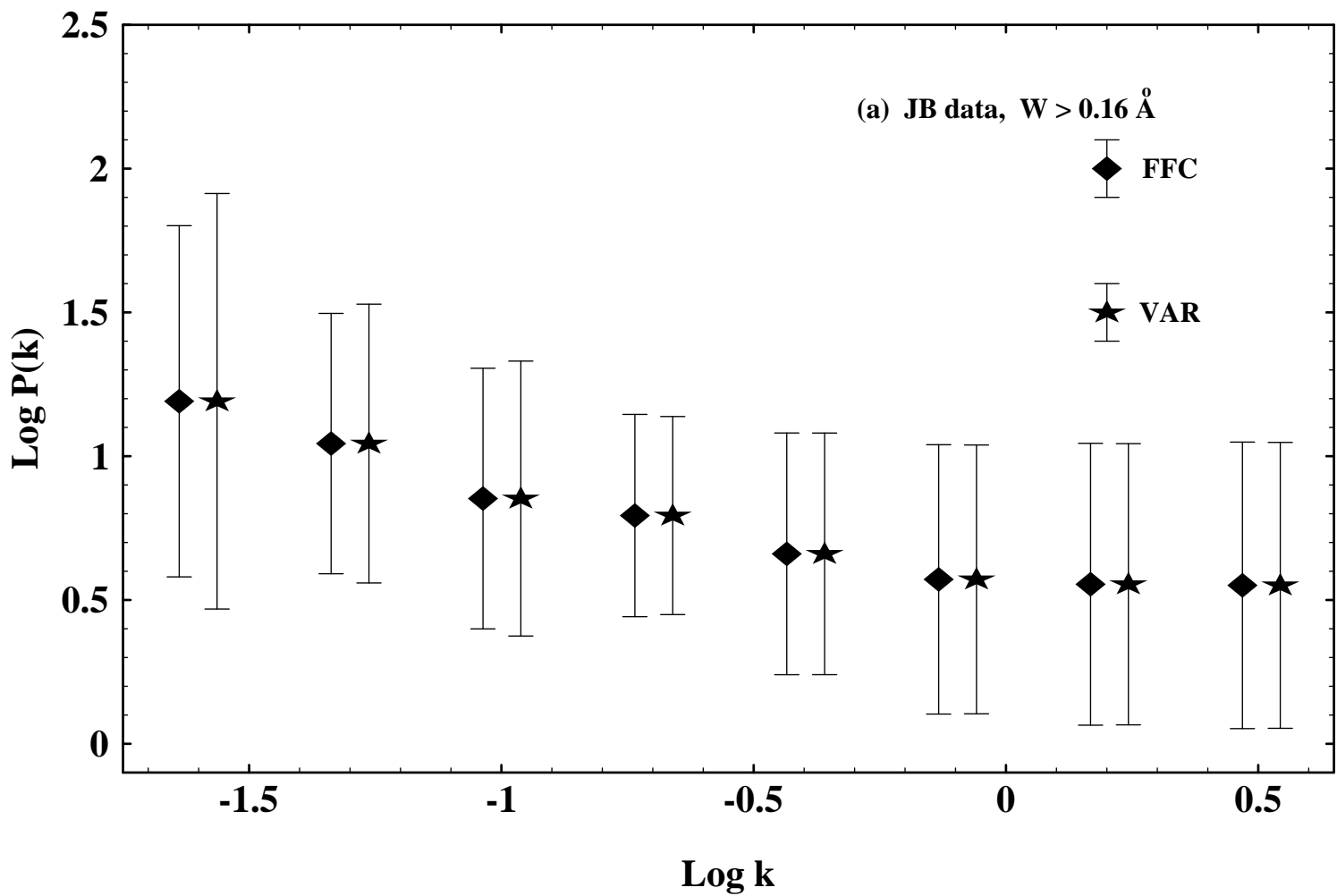

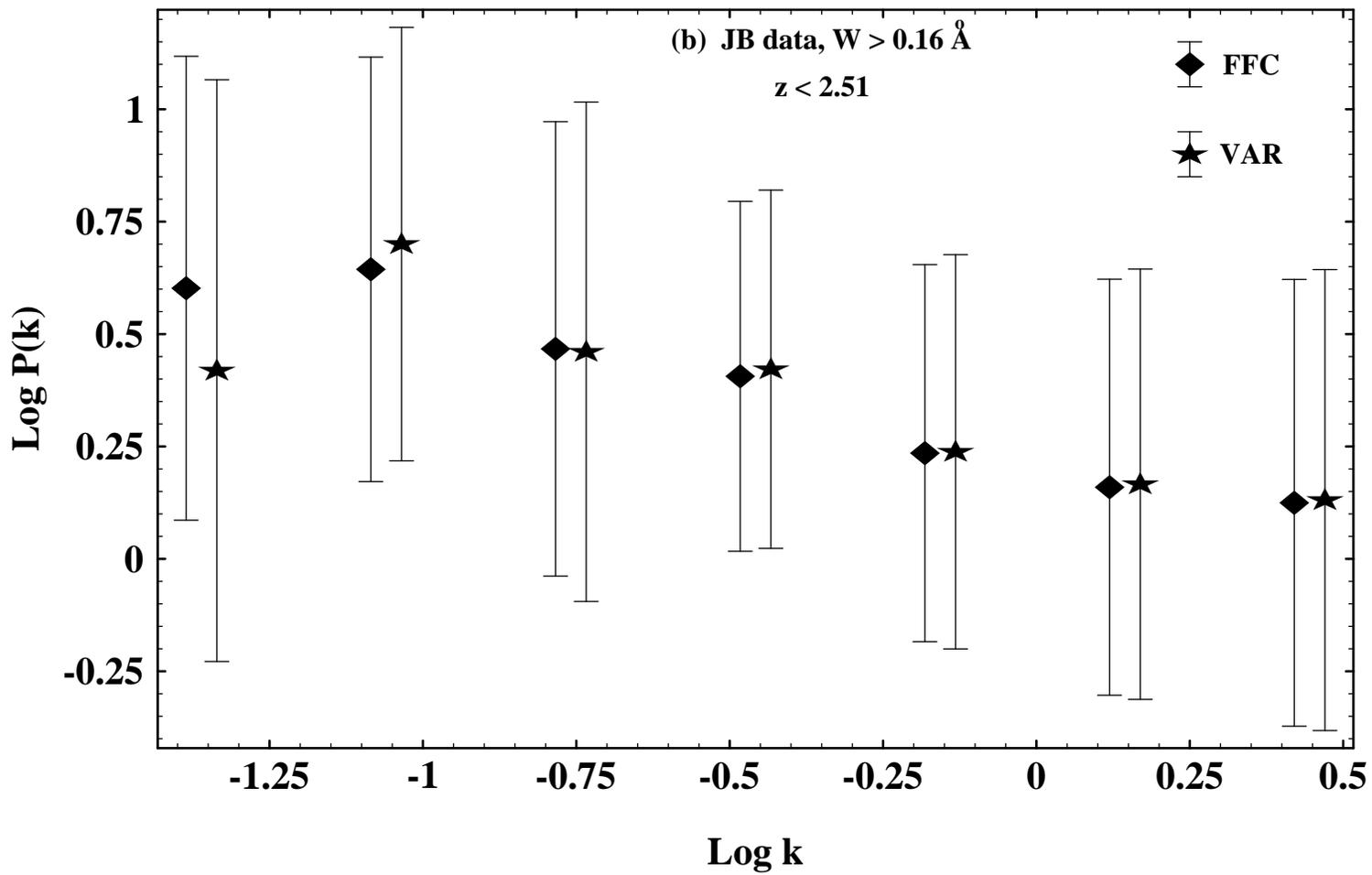

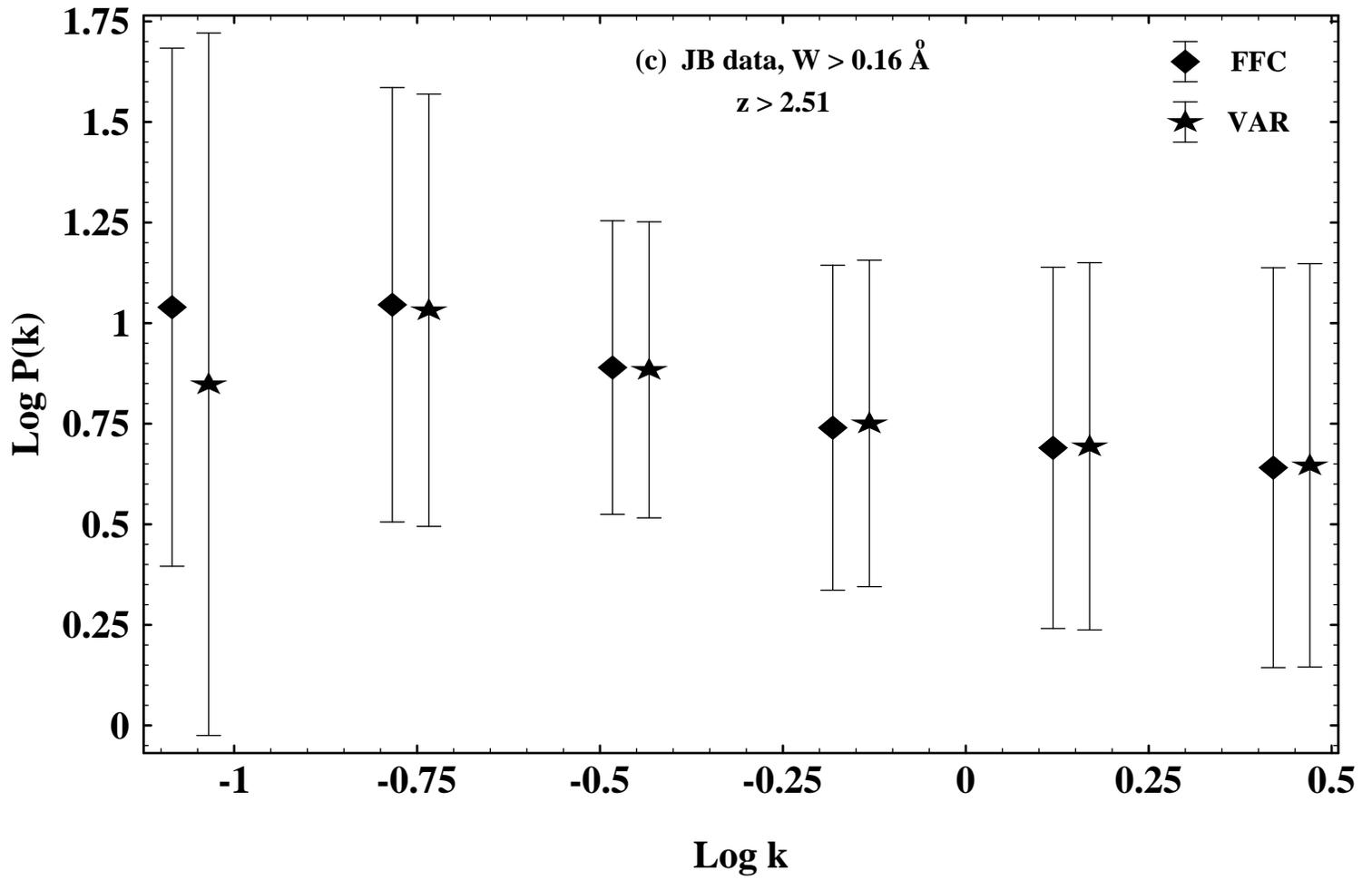

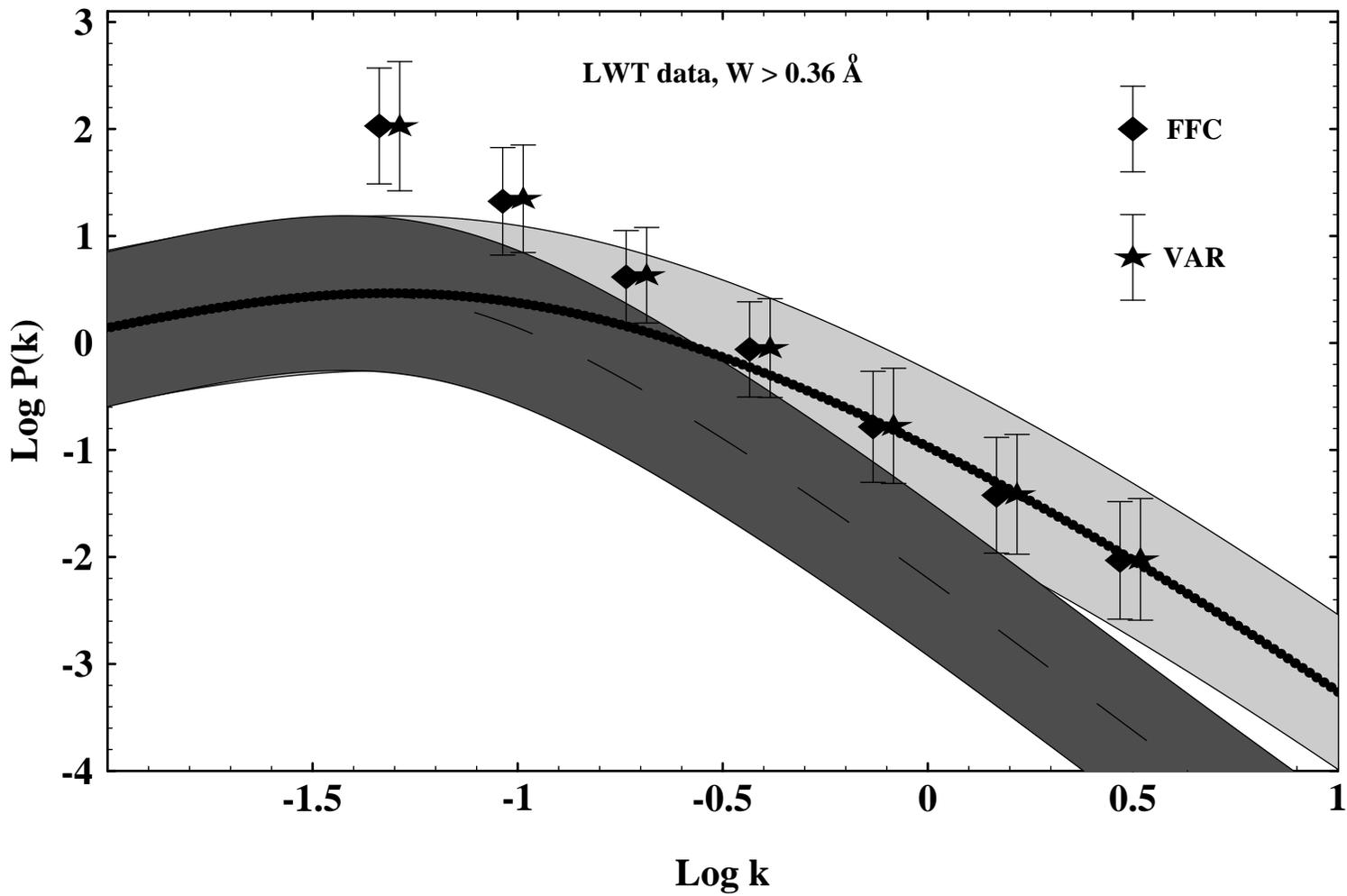

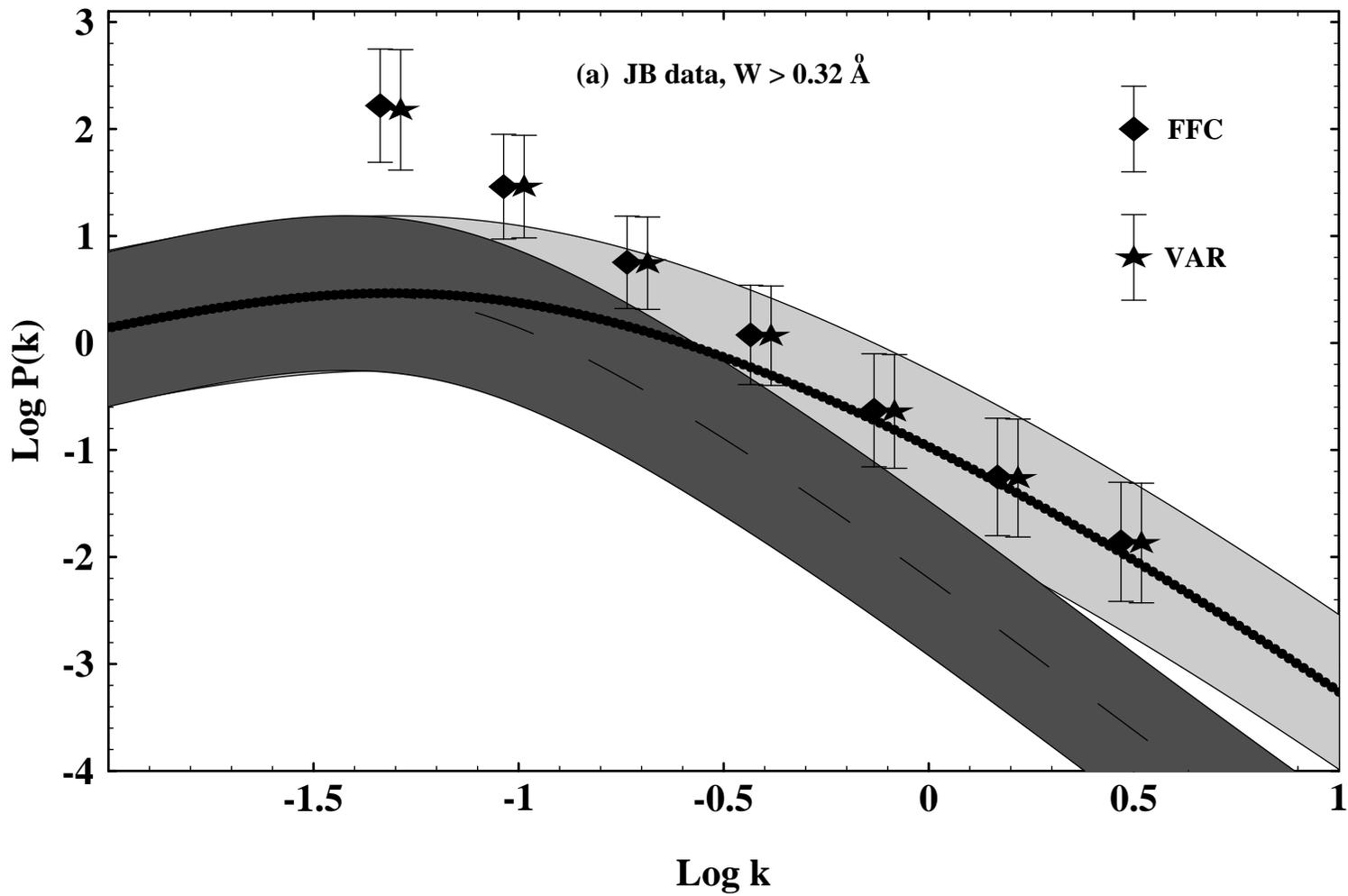

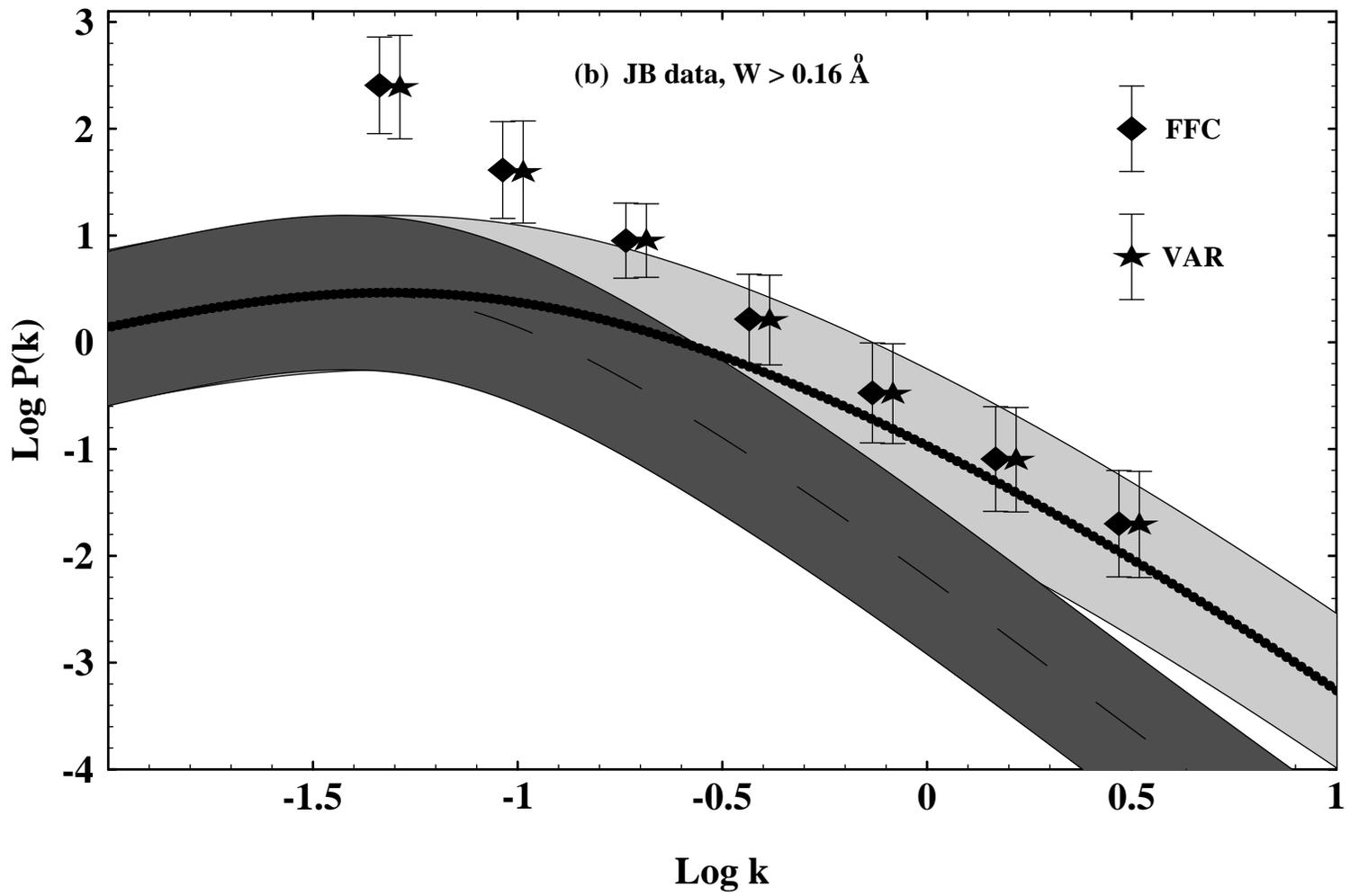